\documentclass[prd,aps,nofootinbib,onecolumn,showpacs,floatfix,10pt]{revtex4}
\usepackage{amsmath,graphicx,color,epsfig}

\def\pp{{\prime\prime}}
\def\vp{\varepsilon}

\begin{document}
\title{\hfill{\small BARI-TH/631-10}\\[0.5em]
$B_s\to D_s(3040)$ form factors and  $B_s$ decays into $D_s(3040)$}
\author{Gang Li $^a$~\footnote{Email:gli@mail.ihep.ac.cn},
 Feng-Lan Shao $^a$~\footnote{Email:shaofl@mail.sdu.edu.cn}
 and Wei Wang $^b$~\footnote{Email:wwang@mail.ihep.ac.cn}
} \affiliation{$^a$ Department of Physics, Qufu Normal University,
Shandong 273165, People's Republic of China \\
 $^b$  Istituto Nazionale di Fisica Nucleare, Sezione di Bari, Bari 70126, Italy  }

\begin{abstract}
Under the assignment of $D_s(3040)$ as a radially excited p-wave
$\bar cs$ state with  $J^P=1^+$, we compute the $B_s\to D_s (3040)$
form factors within the covariant light-front quark model. Two
classification schemes for the p-wave $\bar cs$ meson are adopted.
We also use our results to predict the branching ratios (BRs),
polarization fractions, and angular asymmetries in semileptonic
$B_s\to D_s(3040)\ell\bar\nu(\ell=e,\mu,\tau)$. The BRs are found at
the order of $10^{-3}$ for $\ell=e,\mu$, and $10^{-5}$ for
$\ell=\tau$. We find that the polarization fractions and angular
asymmetries could be useful to pin down the ambiguities in
theoretical prescriptions for  $D_s(3040)$. In addition, we
investigate the nonleptonic $B_s\to D_s(3040)M$ decays under the
factorization method, where $M$ denotes a charged pseudoscalar or a
vector meson. The BRs of $B_s\to D_s(3040)\rho$ and $B_s\to
D_s(3040)D_s^*$ reach the order of $10^{-3}$, while the other
channels are typically smaller by 1-2 orders.
\end{abstract}

\maketitle

\section{Introduction}

The observations of a number of new resonances, such as the narrow
states $D_s(2317)$ in the $D_s^+\pi^0$ final state and $D_s(2460)$
in the $D_s^*\pi^0$ and $D_s\gamma$
channel~\cite{Aubert:2003fg,Besson:2003cp}, have brought great
opportunities for our understanding of nonperturbative QCD in the
charm mass region. They also initiate tremendous efforts on the
exploration of exotic hadrons beyond the naive quark model and
revive great interests in the study of the $ c\bar s$
spectrum~\cite{DeFazio:2009xd}.


In the heavy quark limit, the heavy quark will decouple from the
light degrees of freedom and act as a static color source. Strong
interactions will be independent of the heavy flavor and spin. In
this case, heavy-light mesons, the eigenstates of the QCD Lagrangian
in the heavy quark limit, can be labeled according to the total
angular momentum $s_l$ of the light degrees of freedom. Heavy mesons
with the same $s_l$ but different spin orientations of the heavy
quark fill into one doublet. Accordingly, heavy mesons can be
classified by these doublets, characterized by $s_l$ instead of the
commonly-used quantum numbers, $^{2S+1}L_J$.

The classification scheme based on the heavy quark symmetry applies
to the $D_{sJ}$ mesons. For instance, $D_s(1969)$ and $D_s^*(2112)$
can be arranged into the lowest-lying doublet with
$J^P_{s_l}=(0^-,1^-)_{1/2}$. Among the p-wave $c\bar s$ states,
$D_{s0}(2317)$ and $D_{s1}(2460)$ belong to the doublet with
$J^P_{s_l}=(0^+,1^+)_{1/2}$, and doublet $J^P_{s_l}=(1^+,2^+)_{3/2}$
can be filled by $D_{s1}(2536)$ and $D_{s2}(2573)$. Despite some
controversies over their internal structures, these mesons have been
well established in experiment.

Several other resonant signals have also been observed recently on
the $B$ factories and other facilities. The SELEX collaboration
reported one narrow state $D_{sJ}(2632)$ in
2004~\cite{Evdokimov:2004iy}, which however, has not been confirmed
so far by other experiments. $D_s(2710)$ is announced in $B$ meson
decays by the Belle collaboration~\cite{Abe:2006xm} and also
confirmed by the BaBar results~\cite{Aubert:2009di}. The $D_s(2860)$
is discovered in the $DK$ final state by BaBar~\cite{Aubert:2006mh}.
Recently, the BaBar collaboration has reported one new $D_{sJ}$
state in the study of the $e^+e^-$
annihilations~\cite{Aubert:2009di}. The mass and decay width of this
broad state are
\begin{eqnarray}
 m&=&(3044\pm8^{+30}_{-5}){\rm
 MeV},\;\;\;\Gamma=(239\pm35^{+46}_{-42}){\rm MeV}.
\end{eqnarray}
Its quantum numbers could be $J^P=0^-, 1^+, 2^-...$, taking into
account its signal in $D^*K$ instead of $DK$. The angular analysis
is not available because of the limited statistics.

During the past few years, some progresses have been made in the
classification of these states and understanding of their peculiar
properties in the production or decay processes.
Towards this direction, if the $3^-$ assignment of
$D_s(2710)$~\cite{Colangelo:2006rq,Zhang:2006yj,Chen:2009zt} is
confirmed by the data (see Ref.~\cite{vanBeveren:2009jq} for an
alternative scenario), the $D_s(3040)$ would be unfavored to be
$2^-$ since the mass inversion seems unlikely. On the other hand,
the $1^+$ assignment is favored in several aspects. Firstly, this
explanation is supported by the study of the spectroscopy of excited
$D_s$ mesons~\cite{Di Pierro:2001uu,Matsuki:2006rz,Ebert:2009ua},
though predictions in Refs.~\cite{Close:2006gr,Wang:2007av} are
lower than the experimental value by roughly 200 MeV.
Reference~\cite{Chen:2009zt} investigates these new $D_{sJ}$ mesons
in a semi-classic flux tube model and concludes that $D_s(3040)$ is
compatible with the $1^+$ assignment. The authors of
Ref.~\cite{Colangelo:2010te} have studied the strong decays of
$D_s(3040)$ with four different assignments, and find that the decay
width with $J^P=1^+$ is also consistent with the value given by
BaBar. The strong decays of $D_s(3040)$ are also investigated in the
$^3P_0$ constituent quark model~\cite{Sun:2009tg} and chiral quark
model~\cite{Zhong:2009sk}, respectively.


The verification of the above new states can be via not only the
analysis of their decay properties, which have been widely discussed
in the literature, but also via their productions, which so far have
not been investigated in detail. The production processes of the
above mentioned new states in $B_s$ decays would be an ideal way of
probing their properties. Productions of $D_s(2317)$ and $D_s(2460)$
in the $B_s$ decays have received some theoretical
attentions~\cite{Huang:2004et,Zhao:2006at,Aliev:2006qy,Aliev:2006gk,Li:2009wq}.
On the experimental side, a large number of $B_s$ events will be
collected by the LHCb~\cite{Buchalla:2008jp} and forthcoming Super B
factory experiment~\cite{Aushev:2010bq}. From the theoretical
viewpoint, the degrees of freedom over the $m_b$ scale in $B_s$
decays can be computed in the perturbation theory and the evolution
between the $m_W$ and $m_b$ can be organized using the
renormalization group. The low-energy effects, which are
parameterized into hadronic form factors, will then probe the
structures of the $D_{sJ}$ mesons.

To proceed, we will study the $B_s\to D_{sJ}$ form factors and the production rates of $D_{sJ}$ mesons in
semileptonic and nonleptonic $B_s$ decays in the framework of a
covariant light-front quark model (LFQM)~\cite{Jaus:1999zv}. This
approach is suitable for handling the heavy-light mesons and has
been successfully applied to various
processes~\cite{Cheng:2003sm,Cheng:2004yj,Ke:2009ed,Ke:2009mn,Cheng:2009ms,Wang:2007sxa}.
In particular, we will focus on the $D_s(3040)$ with its assignment
as a $2P$ state. The $B_s\to D_s^*$ transition form factors will be
extracted as a byproduct.

%

This paper is organized as follows. In the next section, an
introduction to the light-front quark model and expressions for the
transition form factors between the $B_s$ and $(D_s^*,D_{s}(3040))$
mesons are presented. In Sec.~III, we compute the form factors under
two different assignments for the $D_s(3040)$. With the form factors
at hand, we will explore in Sec.~IV the semileptonic $B_s$ decays
and the nonleptonic $B_s\to D_s(3040)$ decays under the
factorization assumption. The last section contains our conclusions.

\section{Form factors in the covariant LFQM}

In the effective electroweak Hamiltonian responsible for $B_s \to
D_{sJ}l\bar\nu$~\cite{Buchalla:1995vs}
\begin{eqnarray}
 H_{\rm eff}&=& \frac{G_F}{\sqrt 2} V_{cb} [\bar
 c\gamma_\mu(1-\gamma_5)b][\bar l\gamma^\mu(1-\gamma_5)\nu],
\end{eqnarray}
$G_F$ and $V_{cb}$ are the Fermi constant and the
Cabibbo-Kobayashi-Maskawa (CKM) matrix element, respectively.  The
leptonic part can be directly calculated using perturbation theory.
The residual part contains hadronic effects and  will be
incorporated into the transition form factors as
\begin{eqnarray}
   \langle D_s^*(P^\pp,\vp^{\pp*})|V_\mu|\overline B_s(P^\prime)\rangle &=&
       -\frac{1}{ m_{B_s}+m_{D_s^*}}\,\epsilon_{\mu\nu\alpha \beta}\vp^{\pp*\nu}P^\alpha
    q^\beta  V^{B_sD_s^*}(q^2),    \nonumber\\
\ \ \ \
      \langle D_s^*(P^\pp,\vp^{\pp*})|A_\mu|
    \overline B_s(P^\prime)\rangle &=& i\Big\{
         (m_{B_s}+m_{D_s^*})\vp^{\pp*}_\mu A_1^{B_sD_s^*}(q^2)-\frac{\vp^{\pp*}\cdot P}
         { m_{B_s}+m_{D_s^*}}\,
         P_\mu A_2^{B_sD_s^*}(q^2)    \nonumber \\
    && -2m_{D_s^*}\,{\frac{\vp^{\pp*}\cdot P}{
    q^2}}\,q_\mu\big[A_3^{B_sD_s^*}(q^2)-A_0^{B_sD_s^*}(q^2)\big]\Big\},\nonumber\\
   \langle D_{s1}(P^\pp,\vp^{\pp*})|A_\mu|\overline B_s(P^\prime)\rangle &=&
       -\frac{i}{ m_{B_s}-m_{D_{s1}}}\,\epsilon_{\mu\nu\alpha \beta}\vp^{\pp*\nu}P^\alpha
    q^\beta  A^{B_sD_{s1}}(q^2),    \nonumber \\
\ \ \ \
      \langle D_{s1}(P^\pp,\vp^{\pp*})|V_\mu|
    \overline B_s(P^\prime)\rangle &=& -\Big\{
         (m_{B_s}-m_{D_{s1}})\vp^{\pp*}_\mu V_1^{B_sD_{s1}}(q^2)-\frac{\vp^{\pp*}\cdot P}
         { m_{B_s}-m_{D_{s1}}}\,
         P_\mu V_2^{B_sD_{s1}}(q^2)    \nonumber \\
    && -2m_{D_{s1}}\,{\frac{\vp^{\pp*}\cdot P}{
    q^2}}\,q_\mu\big[V_3^{B_sD_{s1}}(q^2)-V_0^{B_sD_{s1}}(q^2)\big]\Big\},
 \end{eqnarray}
where $P=P^\prime+P^{\prime\prime}$, $q=P^\prime-P^{\prime\prime}$,
and the convention $\epsilon_{0123}=1$ is adopted. The vector and
axial-vector currents are defined as $\bar c\gamma_\mu b$ and $\bar
c\gamma_\mu\gamma_5 b$. We have adopted the $1^+$ assignment,
denoted as $D_{s1}$ hereafter, for the $D_s(3040)$ meson. In the
subsequent analysis two kinds of axial-vectors ($^{2S+1}L_J=^3P_1$
or $^1P_1$) will be considered, and for short we will also
abbreviate them as $^3A$ and $^1A$, respectively. To smear the
singularity at $q^2=0$, we obtain the constraints on the form factors
$A_3^{B_sD_s^*}(0)=A_0^{B_sD_s^*}(0)$ and
$V_3^{B_sD_{s1}}(0)=V_0^{B_sD_{s1}}(0)$, where
\begin{eqnarray}
 A_3^{B_sD_s^*}(q^2)&=&\frac{m_{B_s}+m_{D_s^*}}{ 2m_{D_s^*}}
 A_1^{B_sD_s^*}(q^2)-\frac{m_{B_s}-m_{D_s^*}}{2m_{D_s^*}}\,A_2^{B_sD_s^*}(q^2),\\
 V_3^{B_sD_{s1}}(q^2)&=&\frac{m_{B_s}-m_{D_{s1}}}{ 2m_{D_{s1}}}
 V_1^{B_sD_{s1}}(q^2)-\frac{m_{B_s}+m_{D_{s1}}}{2m_{D_{s1}}}\,V_2^{B_sD_{s1}}(q^2).\label{eq:relation}
 \end{eqnarray}

%

To be specific, we will compute these transition form factors by
considering the general $P\to V$ and $P\to A$ transitions, where $A$
denotes either $^3P_1$ or $^1P_1$ state. The LFQM is based on the
light-front field theory, where the plus component of the gluon
degrees of freedom vanishes in the light-cone gauge.  In this
framework, it is convenient to use the light-front decomposition of
the momentum $P^{\prime}=(P^{\prime -}, P^{\prime +},
P^\prime_\bot)$, with $P^{\prime\pm}=P^{\prime0}\pm P^{\prime3}$, so
that $P^{\prime 2}=P^{\prime +}P^{\prime -}-P^{\prime 2}_\bot$. The
incoming (outgoing) meson has the momentum of
$P^{\prime}=p_1^{\prime}+p_2$ ($P^{\pp}=p_1^{\pp}+p_2$) and mass of
$M^\prime$ ($M^\pp$). The quark and antiquark inside the incoming
(outgoing) meson have the mass $m_1^{\prime(\pp)}$ and $m_2$. Their
momenta are denoted as $p_1^{\prime(\pp)}$ and $p_2$, respectively.
In particular, these momenta can be expressed in terms of the
internal variables $(x_i, p_\bot^\prime)$
 \begin{eqnarray}
 p_{1,2}^{\prime+}=x_{1,2} P^{\prime +},\qquad
 p^\prime_{1,2\bot}=x_{1,2} P^\prime_\bot\pm p^\prime_\bot,
 \end{eqnarray}
with $x_1+x_2=1$. Using these internal variables, one can define
some useful quantities for the incoming meson
\begin{eqnarray}
 M^{\prime2}_0
          &=&(e^\prime_1+e_2)^2=\frac{p^{\prime2}_\bot+m_1^{\prime2}}
                {x_1}+\frac{p^{\prime2}_{\bot}+m_2^2}{x_2},\quad\quad
                \widetilde M^\prime_0=\sqrt{M_0^{\prime2}-(m^\prime_1-m_2)^2},
\nonumber\\
 e^{(\prime)}_i
          &=&\sqrt{m^{(\prime)2}_i+p^{\prime2}_\bot+p^{\prime2}_z},\quad\qquad
 p^\prime_z=\frac{x_2 M^\prime_0}{2}-\frac{m_2^2+p^{\prime2}_\bot}{2 x_2
 M^\prime_0},
 \end{eqnarray}
where $e_i$ can be interpreted as the energy of the quark or the
antiquark, and $M_0^\prime$ is the  kinetic invariant mass of the
meson system. One advantage of these internal quantities is that
they are Lorentz invariant.  The definition of the internal
quantities for the outgoing meson is similar. To formulate the
amplitude for the transition form factor, we also require the
Feynman rules for the meson-quark-antiquark vertices
($i\Gamma^\prime_M$)
\begin{eqnarray}
 i\Gamma_P^\prime &=& H_P'\gamma_5,
\nonumber\\
i\Gamma_V^\prime&=&iH^\prime_{V}[\gamma_\mu-\frac{1}{W^\prime_{V}}(p^\prime_1-p_2)_\mu],\nonumber
            \\
i\Gamma_{^3A}^\prime&=&iH^\prime_{^3A}[\gamma_\mu+\frac{1}{W^\prime_{^3A}}(p^\prime_1-p_2)_\mu]\gamma_5,
\nonumber\\
i\Gamma_{^1A}^\prime&=&iH^\prime_{^1A}[\frac{1}{W^\prime_{^1A}}(p^\prime_1-p_2)_\mu]\gamma_5.
\label{eq:meson-quark-antiquark}
\end{eqnarray}
In the case of the outgoing meson, we shall use $i(\gamma_0
\Gamma^{\prime\dagger}_M\gamma_0)$ for the relevant vertices.

\begin{figure}
\includegraphics[scale=0.5]{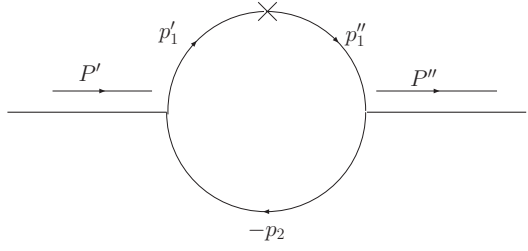}
\caption{Feynman diagram for a transition form factor, where the
cross symbol denotes the electroweak
vertex.}\label{fig:feyn}
\end{figure}

The lowest order contribution to a form factor is depicted in
Fig.~\ref{fig:feyn}, in which the cross label in the diagram denotes
the $V, \ A$ transition vertex. In $B_s$ to $D_{sJ}$ decays,
$p_1'$($p_1''$) is the momentum of the bottom (charm) quark, while
$p_2$ is the momentum of the antiquark.  Physical quantities, e.g.
decay constants and form factors, can be expressed in terms of
Feynman momentum loop integrals which are manifestly covariant. As
an example, we consider the $P\to V$ transition
 \begin{eqnarray}
 {\cal B}^{PV}_\mu=-i^3\frac{N_c}{(2\pi)^4}\int d^4 p^\prime_1
 \frac{H^\prime_P (i H^\pp_V)}{N_1^\prime N_1^\pp N_2} S^{PV}_{\mu\nu}\,\vp^{\pp*\nu},
 \end{eqnarray}
where $N_1^{\prime(\prime\prime)}= p_1^{\prime(\prime\prime)2}
-m_1^{\prime (\prime\prime)2} $ and $N_2=p_2^2-m_2^2$ arise from the
quark propagators and $N_c=3$ is the color factor. The function
$S^{PV}_{\mu\nu}$ is from the Lorentz contraction of the propagators
and the vertex functions
 \begin{eqnarray}
S^{PV}_{\mu\nu} &=&(S^{PV}_{V}-S^{PV}_A)_{\mu\nu}
 \nonumber\\
                &=&{\rm Tr}\left[\left(\gamma_\nu-\frac{1}{W^\pp_V}
                (p_1^\pp-p_2)_\nu\right)
                                 (\not \!p^\pp_1+m_1^\pp)
                                 (\gamma_\mu-\gamma_\mu\gamma_5)
                                 (\not \!p^\prime_1+m_1^\prime)\gamma_5(-\not
                                 \!p_2+m_2)\right].
              \label{eq:BtoV}
 \end{eqnarray}

The form factors will be obtained by evaluating the above formulas.
In practice, we perform the integration over the minus component
within the contour method. If the covariant vertex functions are not
singular when perform the integration, the amplitude will pick up
the singularities in the antiquark propagator.  This manipulation
then leads to
 \begin{eqnarray}
 N_1^{\prime(\pp)}
      &\to&\hat N_1^{\prime(\pp)}=x_1(M^{\prime(\pp)2}-M_0^{\prime(\pp)2}),
 \nonumber\\
 H^{\prime(\pp)}_M
      &\to& h^{\prime(\pp)}_M,
 \nonumber\\
 W^\pp_M
      &\to& w^\pp_M,
 \nonumber\\
\int \frac{d^4p_1^\prime}{N^\prime_1 N^\pp_1 N_2}H^\prime_P H^\pp_V
S^{PV}
      &\to& -i \pi \int \frac{d x_2 d^2p^\prime_\bot}
                             {x_2\hat N^\prime_1
                             \hat N^\pp_1} h^\prime_P h^\pp_V \hat S^{PV},
 \end{eqnarray}
where
 \begin{eqnarray}
 M^{\pp2}_0
          =\frac{p^{\pp2}_\bot+m_1^{\pp2}}
                {x_1}+\frac{p^{\pp2}_{\bot}+m_2^2}{x_2}
 \end{eqnarray}
with $p^\pp_\bot=p^\prime_\bot-x_2\,q_\bot$. The explicit forms of
$h^\prime_M$ and $w^\prime_M$ used in this work are given by~\cite{Jaus:1999zv,Cheng:2003sm}
\begin{eqnarray}
 h^\prime_P&=&h_V'=(M^{\prime2}-M_0^{\prime2})\sqrt{\frac{x_1 x_2}{N_c}}
                    \frac{1}{\sqrt{2}\widetilde M^\prime_0}\varphi^\prime,\label{eqs:s-wave-wa}\nonumber\\
 \sqrt{\frac{2}{3}}h^\prime_{^3A}
                  &=&(M^{\prime2}-M_0^{\prime2})\sqrt{\frac{x_1 x_2}{N_c}}
                    \frac{1}{\sqrt{2}\widetilde M^\prime_0}\frac{\widetilde
                    M^{\prime2}_0}{2\sqrt{3}M^\prime_0}\varphi^\prime_p,\nonumber\\
h^\prime_{^1A}&=&(M^{\prime2}-M_0^{\prime2})\sqrt{\frac{x_1
x_2}{N_c}}
                    \frac{1}{\sqrt{2}\widetilde M^\prime_0}\varphi^\prime_p ,
 \nonumber\\
                     w^\prime_{^3A}&=&\frac{\widetilde
                     M^{\prime2}_0}{m^\prime_1-m_2},\;\;
                    w^\prime_{^1A} =2 ,\label{eqs:wa}
\end{eqnarray}
where $\varphi'$ and $\varphi'_p$ are the light-front wave functions
for the s-wave and p-pave mesons, respectively.

One difference between the conventional
LFQM~\cite{Jaus:1989au,Jaus:1991cy,Cheng:1996if,Choi:2001hg} and the
covariant LFQM lies in the treatment of the constituent quarks. In
the conventional LFQM, the constituent quarks are required to be on
mass shell. Physical quantities can be extracted from the plus
component of the current matrix elements. However, this framework
suffers from the non-covariance problem arising from the missing
zero-mode contributions. This drawback is resolved by including the
so-called $Z$-diagram. In the covariant LFQM the quarks and
antiquarks are off-shell and the physical quantities are written as
the integration over the 4-momentum of the antiquark.  The
conventional LFQM will be recovered after the integration over the
minus component, where the antiquark is set on-shell. The Lorentz
covariance would be lost again as it receives additional
contributions proportional to the light-like four vector
$\tilde\omega=(0,2,{\bf 0_{\perp}})$. This is spurious since the
vector never appears in the definition of the form factors. The
advantage of the covariant LFQM is that we can directly handle the
non-covariant terms instead of compute the $Z$-diagram.
Specifically, the spurious terms can be eliminated by performing the
$p^-$ integration in a proper way and the corresponding rules  are
derived in Refs.~\cite{Jaus:1999zv,Cheng:2003sm}. The above
manipulation results in the expression for $P\to V$ form factors:
\begin{eqnarray}
 g(q^2)&=&-\frac{N_c}{16\pi^3}\int dx_2 d^2 p^\prime_\bot
           \frac{2 h^\prime_P h^\pp_V}{x_2 \hat N^\prime_1 \hat N^\pp_1}
           \Bigg\{x_2 m_1^\prime+x_1 m_2+(m_1^\prime-m_1^\pp)
           \frac{p^\prime_\bot\cdot q_\bot}{q^2}
           +\frac{2}{w^\pp_V}\left[p^{\prime2}_\bot+\frac{(p^\prime_\bot\cdot
            q_\bot)^2}{q^2}\right]
           \Bigg\},
  \nonumber\end{eqnarray}
  \begin{eqnarray}
  f(q^2)&=&\frac{N_c}{16\pi^3}\int dx_2 d^2 p^\prime_\bot
            \frac{ h^\prime_P h^\pp_V}{x_2 \hat N^\prime_1 \hat N^\pp_1}
            \Bigg\{2
            x_1(m_2-m_1^\prime)(M^{\prime2}_0+M^{\pp2}_0)-4 x_1
            m_1^\pp M^{\prime2}_0+2x_2 m_1^\prime q\cdot P
  \nonumber\\
         &&+2 m_2 q^2-2 x_1 m_2
           (M^{\prime2}+M^{\pp2})+2(m_1^\prime-m_2)(m_1^\prime+m_1^\pp)^2
           +8(m_1^\prime-m_2)\left[p^{\prime2}_\bot+\frac{(p^\prime_\bot\cdot
            q_\bot)^2}{q^2}\right]
  \nonumber\\
         &&
           +2(m_1^\prime+m_1^\pp)(q^2+q\cdot
           P)\frac{p^\prime_\bot\cdot q_\bot}{q^2}
           -4\frac{q^2 p^{\prime2}_\bot+(p^\prime_\bot\cdot q_\bot)^2}{q^2 w^\pp_V}
            \bigg[2 x_1 (M^{\prime2}+M^{\prime2}_0)-q^2-q\cdot P
 \nonumber\\
         &&-2(q^2+q\cdot P)\frac{p^\prime_\bot\cdot
            q_\bot}{q^2}-2(m_1^\prime-m_1^\pp)(m_1^\prime-m_2)
            \bigg]\Bigg\},
  \nonumber\end{eqnarray}
  \begin{eqnarray}
 a_+(q^2)&=&\frac{N_c}{16\pi^3}\int dx_2 d^2 p^\prime_\bot
            \frac{2 h^\prime_P h^\pp_V}{x_2 \hat N^\prime_1 \hat N^\pp_1}
            \Bigg\{(x_1-x_2)(x_2 m_1^\prime+x_1 m_2)-[2x_1
            m_2+m_1^\pp+(x_2-x_1)
            m_1^\prime]\frac{p^\prime_\bot\cdot q_\bot}{q^2}
  \nonumber\\
         &&-2\frac{x_2 q^2+p_\bot^\prime\cdot q_\bot}{x_2 q^2
            w^\pp_V}\Big[p^\prime_\bot\cdot p^\pp_\bot+(x_1 m_2+x_2 m_1^\prime)(x_1 m_2-x_2
            m_1^\pp)\Big]\Bigg\},
  \nonumber\end{eqnarray}
  \begin{eqnarray}
 a_-(q^2)&=&\frac{N_c}{16\pi^3}\int dx_2 d^2 p^\prime_\bot
            \frac{ h^\prime_P h^\pp_V}{x_2 \hat N^\prime_1 \hat N^\pp_1}
            \Bigg\{2(2x_1-3)(x_2 m_1^\prime+x_1 m_2)-8(m_1^\prime-m_2)
            \left[\frac{p^{\prime2}_\bot}{q^2}+2\frac{(p^\prime_\bot\cdot q_\bot)^2}{q^4}\right]
 \nonumber\\
         &&-[(14-12 x_1) m_1^\prime-2 m_1^\pp-(8-12 x_1) m_2]\frac{p^\prime_\bot\cdot q_\bot}{q^2}\nonumber\\
         &&+\frac{4}{w^\pp_V}\bigg([M^{\prime2}+M^{\pp2}-q^2+2(m_1^\prime-m_2)(m_1^\pp+m_2)]
                                   (A^{(2)}_3+A^{(2)}_4-A^{(1)}_2)
 \nonumber\\
         &&+Z_2(3 A^{(1)}_2-2A^{(2)}_4-1)+\frac{1}{2}[x_1(q^2+q\cdot P)
            -2 M^{\prime2}-2 p^\prime_\bot\cdot q_\bot -2 m_1^\prime(m_1^\pp+m_2)
 \nonumber\\
         &&-2 m_2(m_1^\prime-m_2)](A^{(1)}_1+A^{(1)}_2-1)
         q\cdot P\Bigg[\frac{p^{\prime2}_\bot}{q^2}
         +\frac{(p^\prime_\bot\cdot q_\bot)^2}{q^4}\Bigg] (4A^{(1)}_2-3)\bigg)
            \Bigg\},\label{eq:expressions-form-factors}
 \end{eqnarray}
with
 \begin{eqnarray}
 V^{B_sD_s^*}(q^2)&=&-(m_{B_s}+m_{D_s^*})\, g(q^2),\quad
 A_1^{B_sD_s^*}(q^2)=-\frac{f(q^2)}{m_{B_s}+m_{D_s^*}},\nonumber\\
 A_2^{B_sD_s^*}(q^2)&=&(m_{B_s}+m_{D_s^*})\, a_+(q^2),\quad
 A_3^{B_sD_s^*}(q^2)-A_0^{B_sD_s^*}(q^2)=\frac{q^2}{2 m_{D_s^*}}\,
 a_-(q^2).\label{eq:relation-vector}
 \end{eqnarray}
The functions $Z_2$, $A^{(1)}_1$, $A^{(1)}_2$, $A^{(2)}_3$, and
$A^{(2)}_4$ are given in the Appendix.

The computation of $B_s\to D_{s}(3040)$ transition is analogous, and
in fact, the $B\to ^3A$ form factors are related to the $B\to V$
ones. In particular, the auxiliary form factors are given as
 \begin{eqnarray}
 \ell^{^3A}(q^2)&=&f(q^2) \,\,\,{\rm with}\,\,\,
                         (m_1^\pp\to -m_1^\pp,\,h^\pp_V\to h^\pp_{^3A},
                         \,w^\pp_V\to w^\pp_{^3A}),
 \nonumber\\
 q^{^3A}(q^2)&=&g(q^2) \,\,\,{\rm with}\,\,\,
                         (m_1^\pp\to -m_1^\pp,\,h^\pp_V\to h^\pp_{^3A},
                         \,w^\pp_V\to  w^\pp_{^3A}),
  \nonumber\\
 c_\pm^{^3A}(q^2)&=&a_\pm(q^2) \,\,\,{\rm with}\,\,\,
                          (m_1^\pp\to -m_1^\pp,\,h^\pp_V\to h^\pp_{^3A},
                          \,w^\pp_V\to  w^\pp_{^3A}),
                          \label{eq:BtoAformfactor}
 \end{eqnarray}
where the replacement of $m_1^\pp\to -m_1^\pp$ does not apply to
$m_1^\pp$ in $w^\pp$ and $h^\pp$, as it arises from the propagators
and quark-antiquark-meson coupling vertex
\begin{eqnarray}
S^{P^3A}_{\mu\nu} &=&(S^{P^3A}_{V}-S^{P^3A}_A)_{\mu\nu}
 \nonumber\\
                &=&{\rm
                Tr}\left[\left(\gamma_\nu-\frac{1}{W^\pp_{^3A}}
                (p_1^\pp-p_2)_\nu\right)\gamma_5
                                 (\not \!p^\pp_1+m_1^\pp)
                                 (\gamma_\mu-\gamma_\mu\gamma_5)
                                 (\not \!p^\prime_1+m_1^\prime)\gamma_5(-\not
                                 \!p_2+m_2)\right]\nonumber \\
                &=&-S^{PV}_{\mu\nu}|_{W_V''\to W_{^3A}'', m_1''\to
                -m_1''}.
\end{eqnarray}
Then, the $B_s\to D_s(^3A)$  form factors can be expressed by
 \begin{eqnarray}
 A^{B_sD_{s1}(^3A)}(q^2)&=&-(m_{B_s}-m_{D_{s1}(^3A)})\, q^{^3A}(q^2),\quad
 V^{B_sD_{s1}(^3A)}_1(q^2)=-\frac{\ell^{^3A}(q^2)}{m_{B_s}-m_{D_{s1}(^3A)}},
 \nonumber\\
 V^{B_sD_{s1}(^3A)}_2(q^2)&=&(m_{B_s}-m_{D_{s1}(^3A)})\, c_+^{^3A}(q^2),\;\;
 V_3^{B_sD_{s1}(^3A)}(q^2)-V_0^{B_sD_{s1}(^3A)}(q^2)=\frac{q^2}{2 m_{D_{s1}(^3A)}}\,
 c_-^{^3A}(q^2).
 \end{eqnarray}
The analysis is similar for the $^1A$ meson but only
the $w_{^1A}''$ terms contribute in the transition form factors,
stemming from the structures of meson-quark-antiquark coupling vertices shown in
Eq.~\eqref{eq:meson-quark-antiquark}.

Before closing this section, it is worth mentioning that the
identification of $D_s(3040)$ with either $^3P_1$ or $^1P_1$
configuration may be inappropriate. As discussed above, in the heavy
quark limit the QCD Lagrangian is invariant under the heavy flavor
and spin rotation. Thus, the heavy quark will decouple from the
remanent part. One direct consequence is that the heavy mesons are
not labeled by the quantum number $^{2S+1}L_J$ but by the total
angular momentum $s_l$ of the light quark degrees of freedom. Finite
heavy quark mass corrections will break  this symmetry, and mesons
with the same spin-parity $J^P$ will mix with each other. The
relation between these two bases is
\begin{eqnarray}
 |P^{3/2}_1\rangle &=& \sqrt{\frac{2}{3}} |^1P_1\rangle +\sqrt{\frac{1}{3}}
 |^3P_1\rangle,\nonumber\\
 |P^{1/2}_1\rangle &=& \sqrt{\frac{1}{3}} |^1P_1\rangle -\sqrt{\frac{2}{3}}
 |^3P_1\rangle.
\end{eqnarray}
As follows, we will compute the $B_s\to D_s(3040)$ form factors and
make predictions for physical observables in the relevant $B_s$
decays under these two schemes.

\section{Numerical results}

Expressions for form factors shown in
Eq.~\eqref{eq:expressions-form-factors} involve the quark masses,
hadron masses and the light front wave functions (LFWFs) $\varphi'$
and $\varphi_p'$. The latter should be obtained by solving the
relativistic Schr\"odinger equation. However, in practice a
phenomenological wave function to describe the hadronic structure is
preferred. In this work, we will use the simple Gaussian-type wave
function which has been extensively adopted in the
literature~\cite{Cheng:2003sm,Cheng:2004yj}
\begin{eqnarray}
 \varphi^\prime
    &=&\varphi^\prime(x_2,p^\prime_\perp)
             =4 \left({\frac{\pi}{{\beta^{\prime2}}}}\right)^{3/4}
               \sqrt{{\frac{dp^\prime_z}{{dx_2}}}}~{\rm exp}
               \left(-{\frac{p^{\prime2}_z+p^{\prime2}_\bot}{{2 \beta^{\prime2}}}}\right),%
\quad\qquad
         \frac{dp^\prime_z}{dx_2}=\frac{e^\prime_1 e_2}{x_1 x_2
         M^\prime_0}.
 \label{eq:wavefn}
\end{eqnarray}
Similarly for the $D_s(3040)$, as one of the $2P$ states,
the wave function is given as
\begin{eqnarray}
 \varphi^\prime_p
    &=&\varphi^\prime_p(x_2,p^\prime_\perp)
             =4\sqrt
 {\frac{2}{3}}\left(\frac{\pi}{\beta^{\prime2}}\right)^{3/4} \sqrt{\frac{2}{{\beta^{\prime2}}}}
 \sqrt{\frac{dp_z}{dx_2}}{\rm exp}\left(-\frac{p_{\bot}^2+p_z^2}{2\beta^{\prime2}}\right)
 \times\left(\frac{p_{\bot}^2+p_z^2}{\beta^{\prime2}}-\frac{3}{2}\right).
\end{eqnarray}

The constituent mass for a heavy quark is usually believed close to
the current quark mass. Running quark masses for a charm and bottom
quark in the $\overline {\rm MS}$ renormalization scheme are
constrained as~\cite{Amsler:2008zzb}
\begin{eqnarray}
 m_b=(4.19^{+0.18}_{-0.06}) {\rm GeV},\;\;\; m_c=(1.27^{+0.07}_{-0.09}){\rm GeV},
\end{eqnarray}
while the corresponding pole masses (with one-loop anomalous dimension) are
\begin{eqnarray}
 m_b=(4.65^{+0.20}_{-0.07}) {\rm GeV},\;\;\; m_c=(1.56^{+0.09}_{-0.11}){\rm
 GeV} \ ,
\end{eqnarray}
with $\alpha_s(m_Z=91.19 {\rm GeV})=0.120$~\cite{Amsler:2008zzb} in
the renormalization  group evolution equation. The quark mass
extracted from the $\Upsilon(1S)$ is
\begin{eqnarray}
 m_b=(4.67^{+0.18}_{-0.06}) {\rm GeV}.
\end{eqnarray}
In a phenomenological approach such as the quark model used in this work,
the constituent quark mass is chosen close to the above results.

As well known, the constituent mass of a light quark largely deviates
from its current mass due to non-negligible contributions from gluon
degrees of freedom.
In this case, only phenomenological values can be employed, and in
principle, they can be constrained within a global fit under a
specific framework. In this work, we will employ the values which
are consistent with the previous studies in the covariant LFQM. In
particular, the same values as in Ref.~\cite{Cheng:2003sm} will be
adopted
\begin{eqnarray}
 m_c=(1.4\pm0.1){\rm GeV}, \;\;\; m_b=(4.64\pm 0.2){\rm GeV},\;\;\; m_s=(0.37\pm0.05){\rm GeV},\label{eq:quarkmasses}
\end{eqnarray}
where a comprehensive study on $D,B$ transition form factors has
been performed and many predictions are found in accordance with the
experimental data. It is worth mentioning that
Ref.~\cite{Cheng:2003sm} only gives the central value for the quark
masses. But here we also include sizable uncertainties to estimate
the sensitivity. The above values in Eq.~\eqref{eq:quarkmasses} are
also consistent with the results derived from the variational
principle in the conventional LFQM~\cite{Choi:1999vu}.


The shape parameter $\beta'$, which describes the momentum
distribution, is fixed by the meson's decay constant, of which the
analytic expression is derived as
\begin{eqnarray}
 f_P&=&\frac{N_c}{16\pi^3}\int
 dx_2d^2p'_\perp\frac{h_P'}{x_1x_2(M^{\prime2}-M_0^{\prime2})}4(m_1'x_2+m_2x_1'),\label{eq:Pdecayconstantusual}\nonumber\\
 f_V&=&\frac{N_c}{M'4\pi^3}\int
 dx_2d^2p'_\perp\frac{h_V'}{x_1x_2(M^{\prime2}-M_0^{\prime2})}
  \times\left[m_1m_2-m_1^2+x_1M^{\prime2}_0-p_\perp^{\prime2}+\frac{p_{\perp}^{\prime2}}{w_V}(m_2+m_1)\right].\label{eq:Vdecayconstantusual}
\end{eqnarray}
For the $B_s$ meson decay constant, we use the latest Lattice QCD
result~\cite{Gamiz:2009ku}:
\begin{eqnarray}
f_{B_s}=(231\pm 15){\rm MeV}.
\end{eqnarray}
The $D_s$ decay constant can be measured by the semileptonic $D_s\to
\mu\bar\nu_\mu$ decays, for which a number of measurements are
available. The averaged value by the HFAG at $1\sigma$ level will be
used in this work~\cite{HFAG}
\begin{eqnarray}
 f_{D_s}&=&(256.9\pm6.8) {\rm MeV}.
\end{eqnarray}
At present, no direct measurement of the $D_s^*$ decay constant is
available. Fortunately, the heavy quark symmetry  (HQS) implies that
this quantity can be related to  $f_{D_s}$, i.e.
\begin{eqnarray}
 f_{D^*_s}= \sqrt{\frac{m_{D_s}}{m_{D^*_s}}} f_{D_s}=(248.0\pm6.6) {\rm MeV}.
\end{eqnarray}
Symmetry breaking effects emerge as power corrections in
$\Lambda_{\rm QCD}/m_Q$  (here $\Lambda_{\rm QCD}$ is the hadronic
scale while $m_Q$ can be chosen as $m_c$ or $m_{D_s,D_s^*}$), of
which the value can amount to $20\%$. In the present work, we will
use the value derived from the HQS relation and neglect the power
corrections. Using the above decay constants, we fix the shape
parameters in the LFWFs as
\begin{eqnarray}
 \beta_{B_s}=(0.6224\pm0.0339){\rm GeV},\;\;\;\beta_{D_s^*}=(0.421\pm0.007){\rm GeV}.
\end{eqnarray}

With the instantaneous Bethe-Salpeter method, the decay constant for
radially excited $1^+$ states has been studied by
Ref.~\cite{Wang:2007av}, i.e.
\begin{eqnarray}
 f_{D_s(2^3P_1)}&=& 204 {\rm MeV},\;\;\; f_{D_s(2^1P_1)}=50 {\rm MeV},
\end{eqnarray}
but no uncertainties can be found in Ref.~\cite{Wang:2007av}. With the value for $f_{D_s(2^3P_1)}$ and
identifying $D_s(3040)$ as one $2^3P_1$ state, we find that the
shape parameter is close to $0.3$ GeV. As a first step to proceed, we will employ this
value with an uncertainty as follows,
\begin{eqnarray}
\beta_{D_s(3040)}=(0.300\pm0.015) {\rm GeV}.
\end{eqnarray}
This choice corresponds to the decay constant
\begin{eqnarray}
 f_{D_s(2^3P_1)}&=& (198^{+20}_{-21}){\rm MeV},\;\;\; f_{D_s(2^1P_1)}=(58\pm 5) {\rm MeV}.
\end{eqnarray}

For later convenience, the decay constants of $D^-$ extracted from
the experimental data of $D^-\to \mu^-\bar\nu$, and $D^{*-}$ from
HQS are also listed:
\begin{eqnarray}
 f_{D}&=&(216.6\pm17.2) {\rm MeV},\;\;\; f_{D^*}= \sqrt{\frac{m_{D}}{m_{D^*}}} f_{D}
 =(209.1\pm 16.8) {\rm MeV}.
\end{eqnarray}
Decay constants of the light mesons are obtained by making use of
the experimental data of $P\to l\bar\nu$ and $\tau\to V^-\nu_\tau$
decays~\cite{Amsler:2008zzb}
\begin{eqnarray}
 f_\pi=0.131 {\rm GeV},\;\;\; f_K=0.16 {\rm GeV},\;\;
 f_\rho=0.2212 {\rm GeV},\;\;\; f_{K^*}=0.2068 {\rm GeV}.
\end{eqnarray}

It is worth pointing out that because of the condition $q^+=0$
imposed in the course of calculation, form factors are known only at
space-like momentum transfer $q^2=-q^2_\bot\leq 0$. On the other
hand, only time-like form factors are relevant in physical decay
processes. Namely, in the exclusive non-leptonic decays, the form
factor at maximally recoiling ($q^2\simeq 0$) is required; while the
$q^2$-dependent behavior in the full $q^2>0$ region is needed in
semileptonic $\bar B_s$ decays. In order to have the behavior in the
whole $q^2$ region, we adopt the parametrization
\begin{eqnarray}
 F(q^2)=\frac{F(0)}{1-aq^2/m_{B_s}^2+b(q^2/m_{B_s}^2)^2},
\end{eqnarray}
where $F$ denotes any generic form factors $V,A_0,A_1,A_2$ for
$B_s\to D_s^*$, or $A,V_0,V_1,V_2$ for $B_s\to D_{s1}$. These
parameters $(F,a,b)$ will be fitted in the space-like region ($-20
{\rm GeV}^2<q^2<0$). Then, the form factors will be extrapolated to
the time-like region.

\begin{figure}
\includegraphics[scale=0.8]{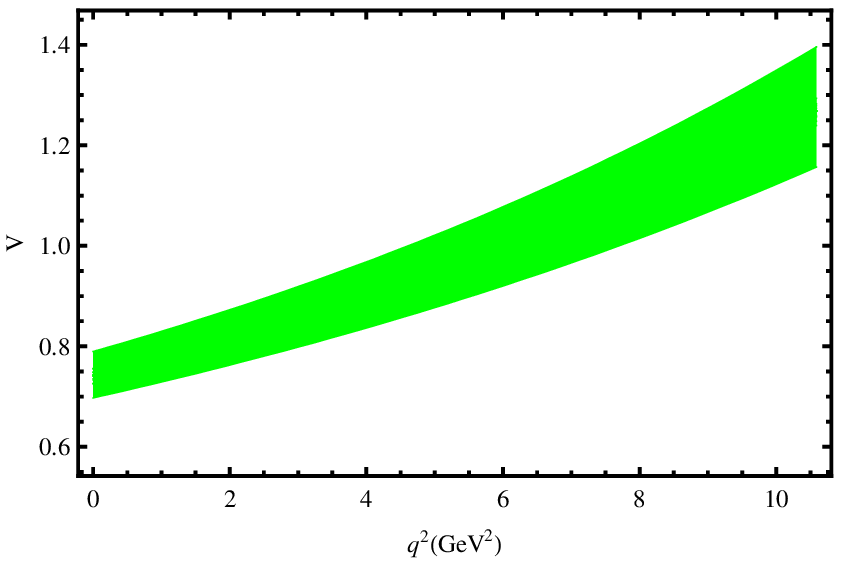}
\includegraphics[scale=0.8]{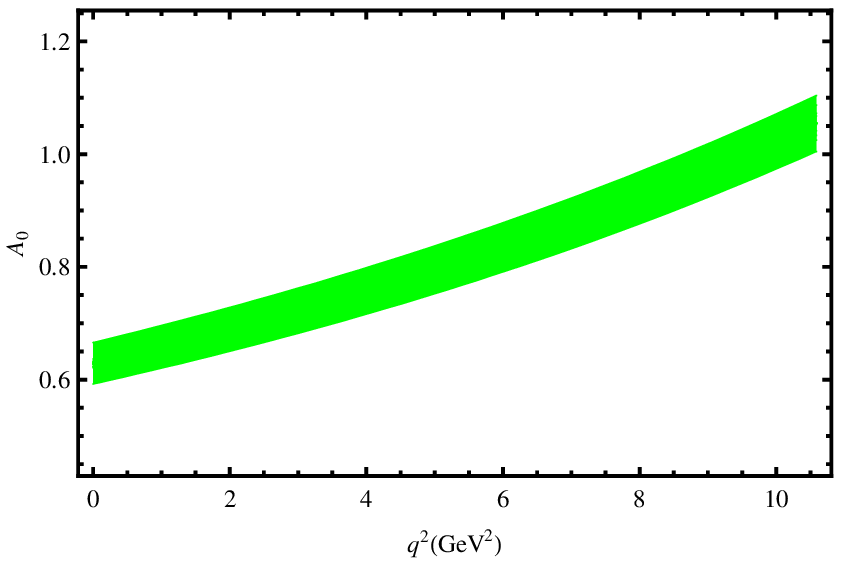}
\includegraphics[scale=0.8]{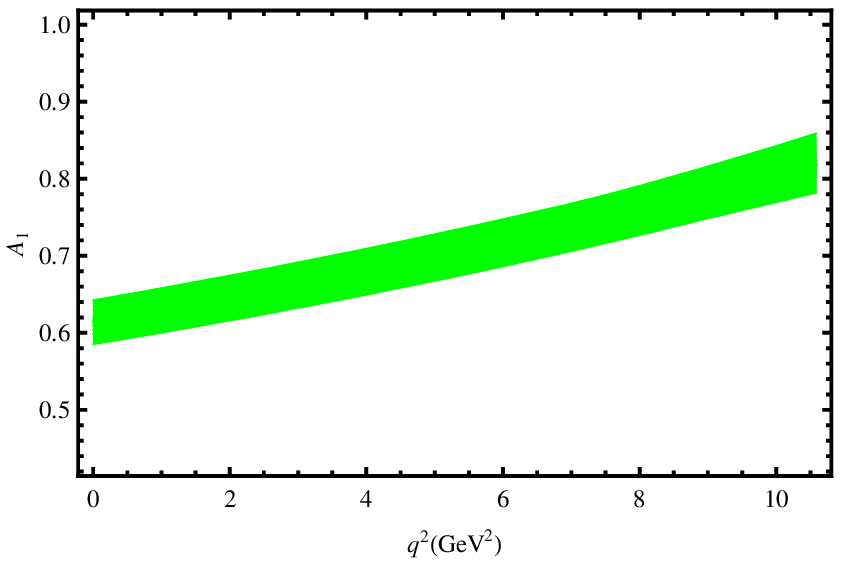}
\includegraphics[scale=0.8]{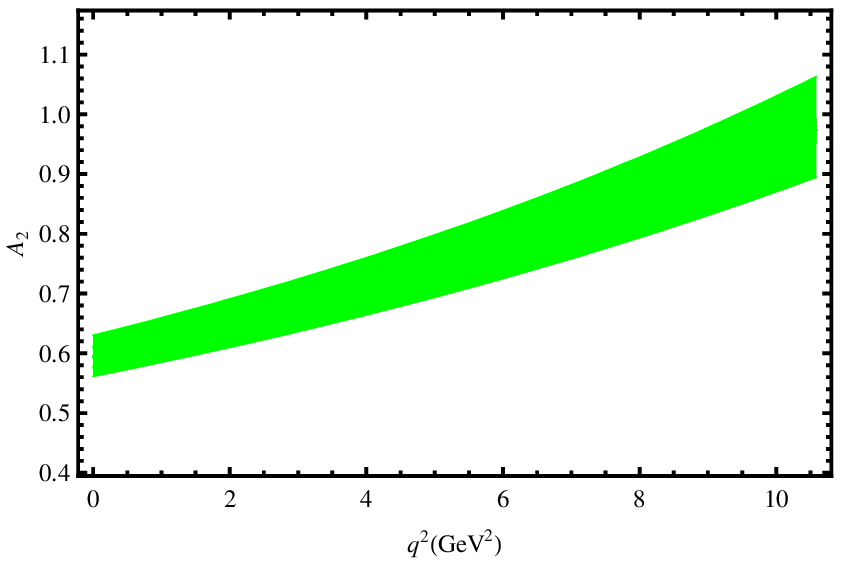}
\caption{The $q^2$-dependence of the $B_s\to D_s^*$ form
factors}\label{fig:q2-Dstar}
\end{figure}

With the above input parameters, our results for the form factors
are collected in table~\ref{tab:form-factors}, where the
uncertainties are from the shape parameters of the LFWFs and the
constituent quark masses. Our strategy for the uncertainties is as
follows. With the central value for quark masses, we first fix shape
parameters so that  these values can reproduce the adopted decay
constants. Secondly, we will also compute the errors caused by quark
masses but keeping the shape parameters unchanged. The final
uncertainties for the form factors and physical observables will be
added in quadrature. We plot the $q^2$-dependence of the form
factors in Fig.~\ref{fig:q2-Dstar}$\sim$
\ref{fig:q2-Ds(3040)A-3over2} for $B_s\to D_s^*$, $D_{s1}(^3P_1)$,
$D_{s1}(^1P_1)$, $D_{s1}(P^{1/2}_1)$, and $D_{s1}(P^{3/2}_1)$,
respectively.

\begin{figure}
\includegraphics[scale=0.8]{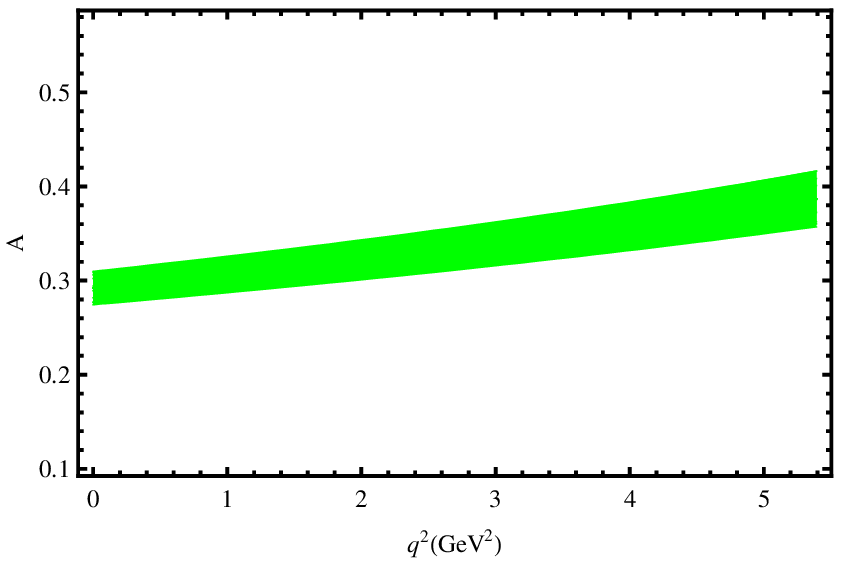}
\includegraphics[scale=0.8]{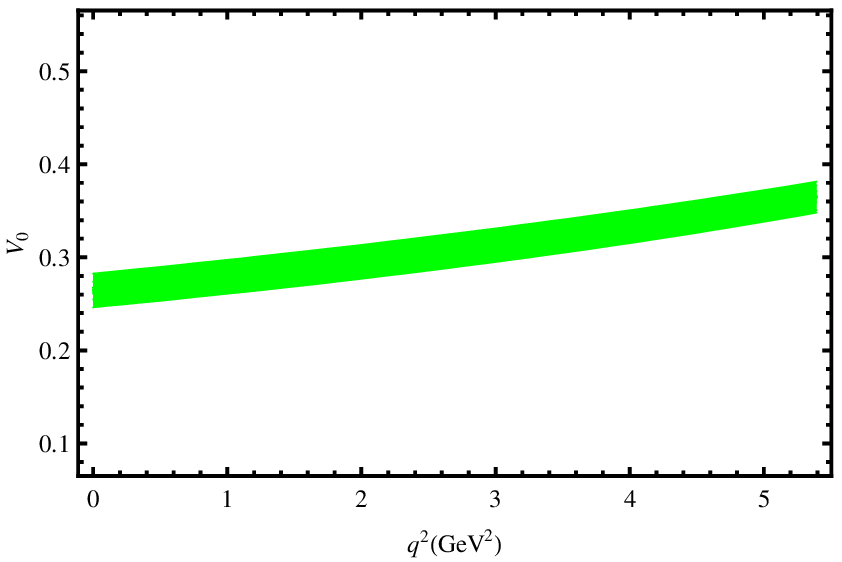}
\includegraphics[scale=0.8]{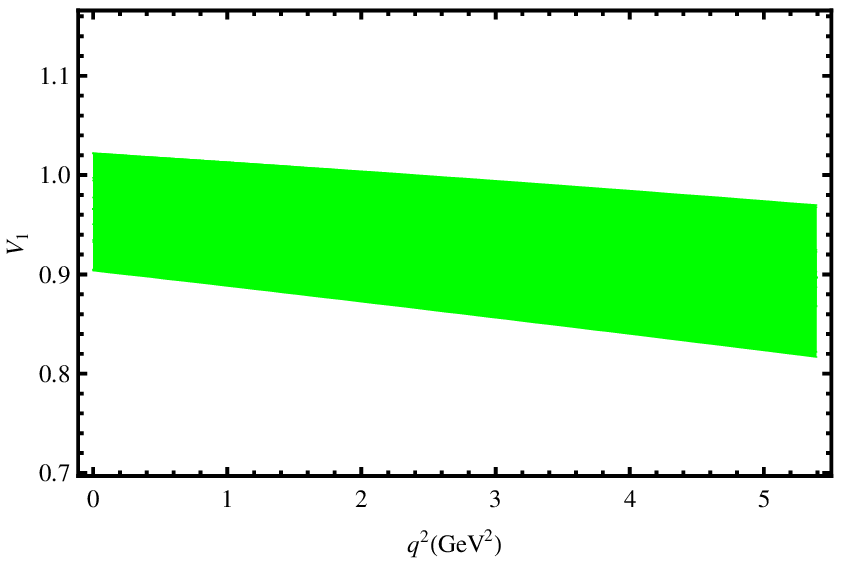}
\includegraphics[scale=0.8]{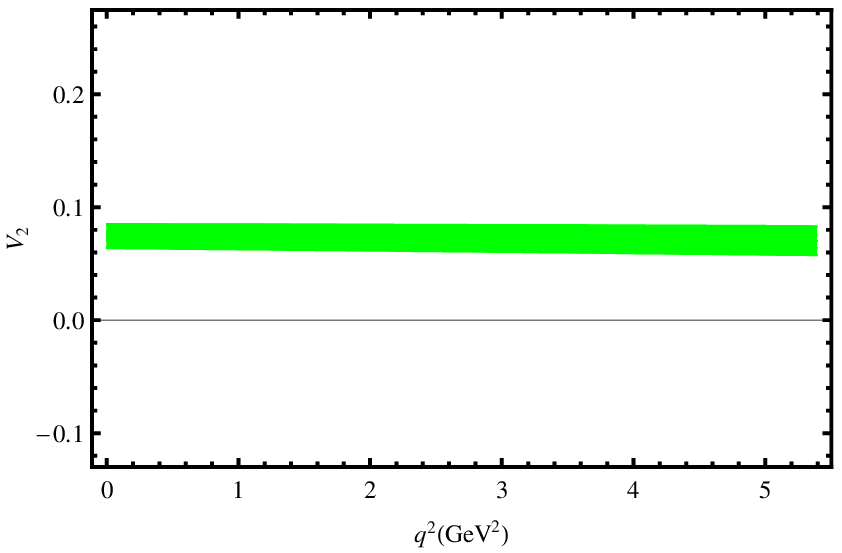}
\caption{The $q^2$-dependence of the $B_s\to D_s(3040)$ form factors
with $^{2S+1}L_J=^3P_1$}\label{fig:q2-Ds(3040)3A}
\end{figure}

\begin{figure}
\includegraphics[scale=0.8]{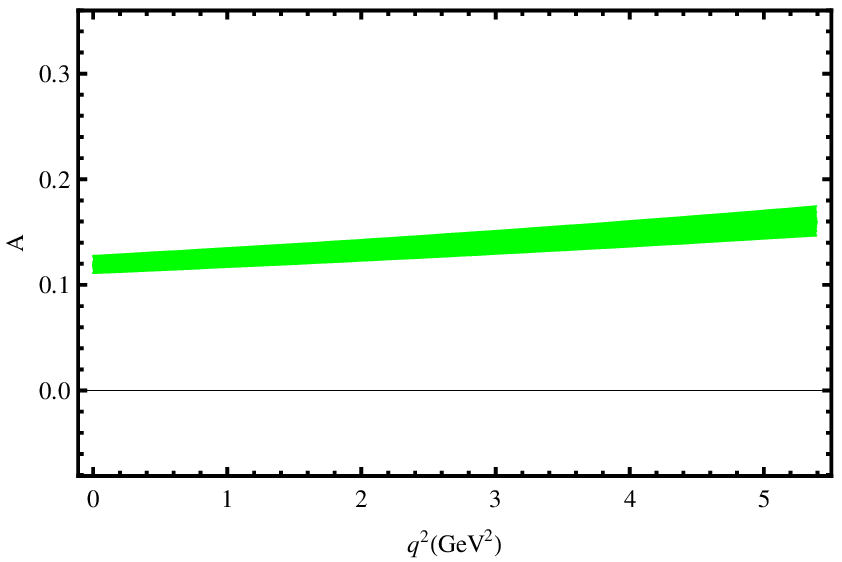}
\includegraphics[scale=0.8]{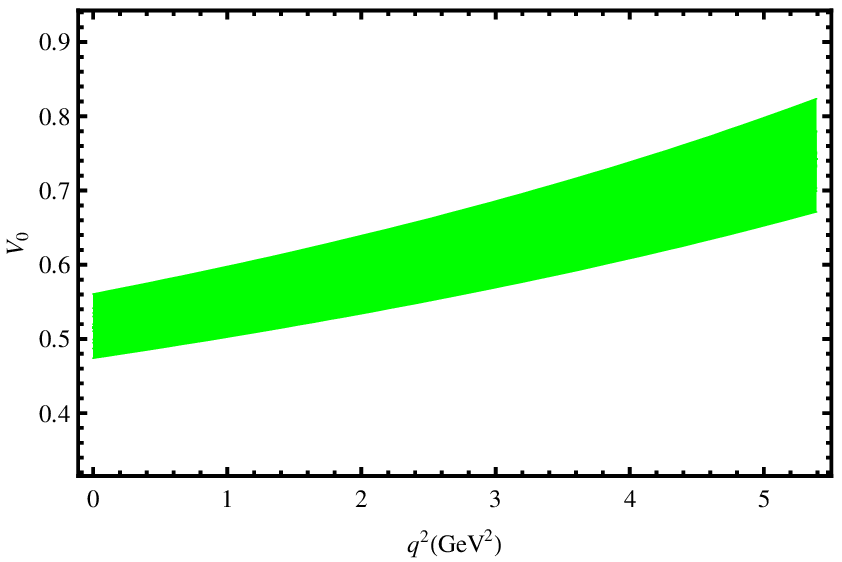}
\includegraphics[scale=0.8]{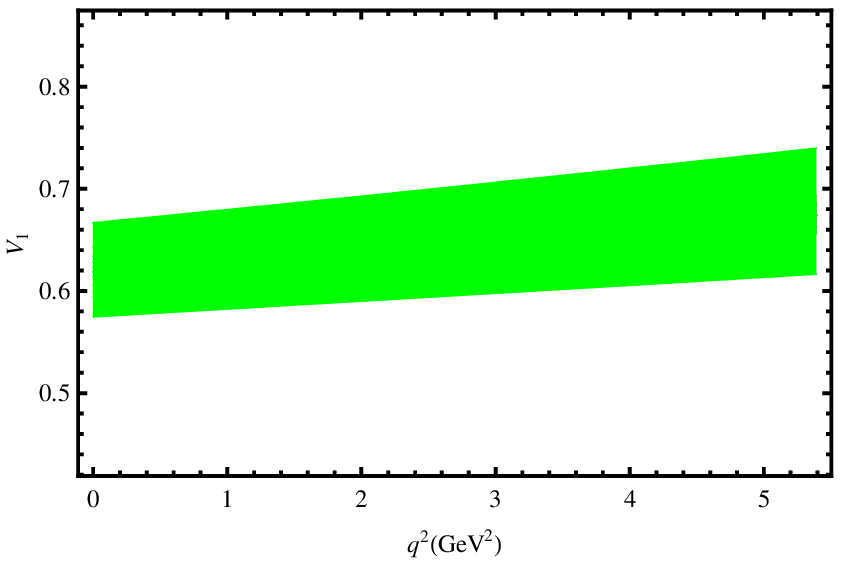}
\includegraphics[scale=0.8]{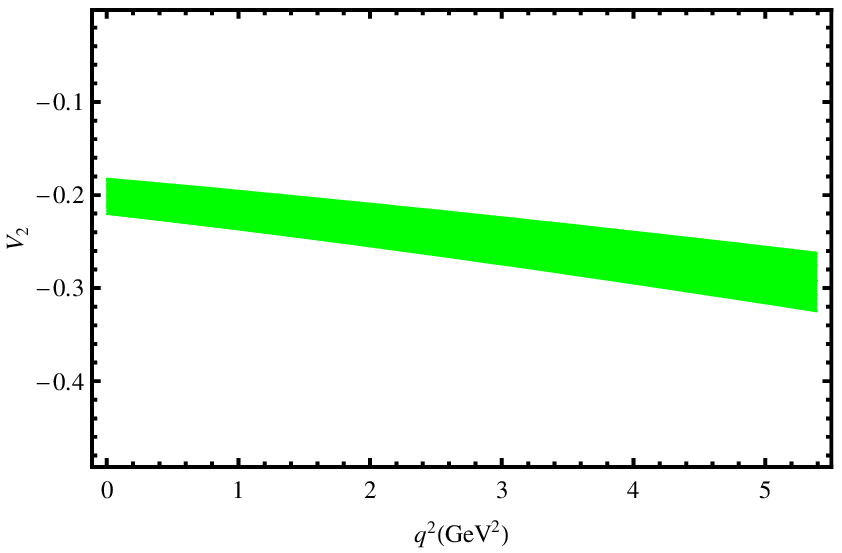}
\caption{The $q^2$-dependence of the $B_s\to D_s(3040)$ form factors
with $^{2S+1}L_J=^1P_1$}\label{fig:q2-Ds(3040)1A}
\end{figure}

\begin{figure}
\includegraphics[scale=0.8]{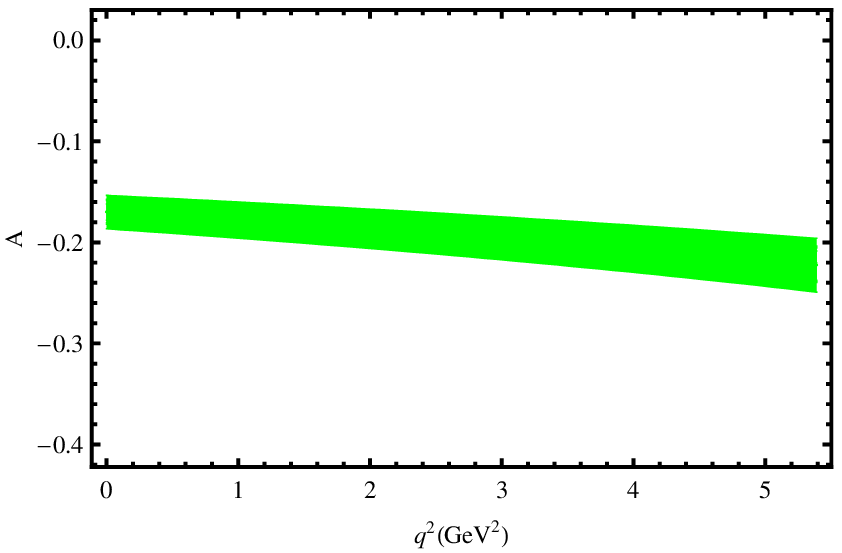}
\includegraphics[scale=0.8]{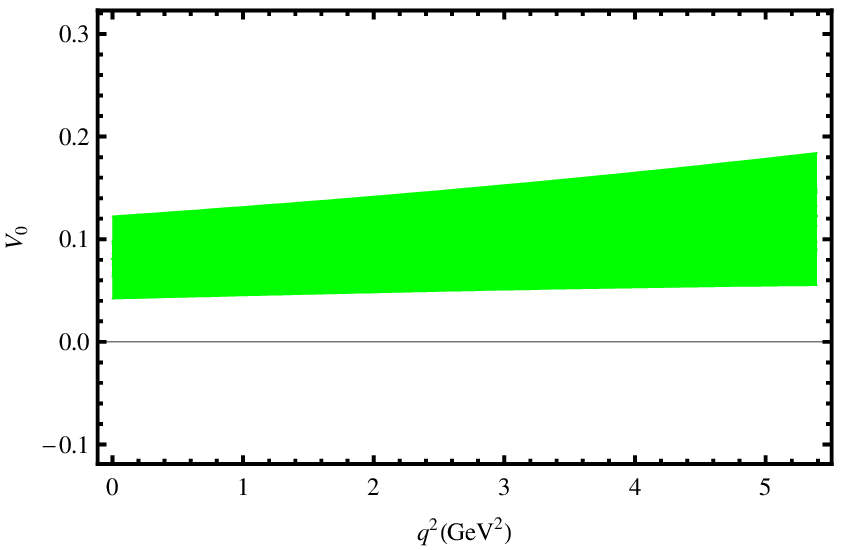}
\includegraphics[scale=0.8]{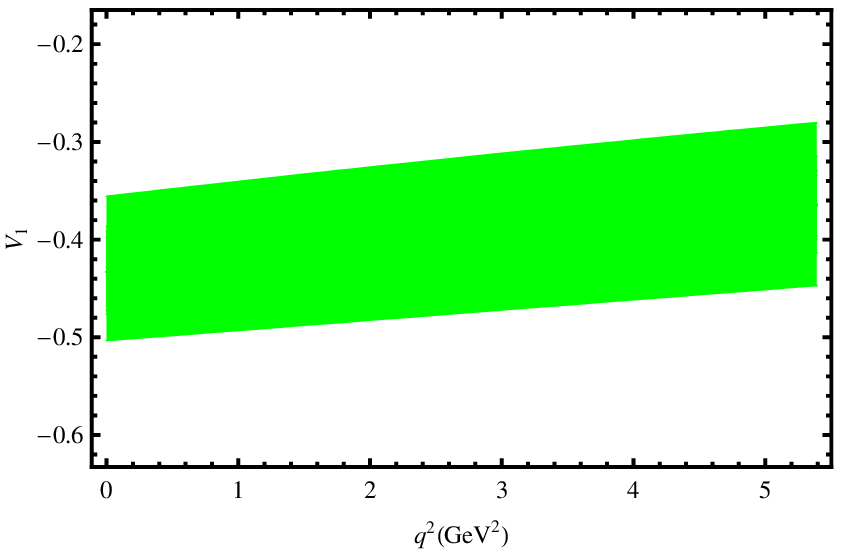}
\includegraphics[scale=0.8]{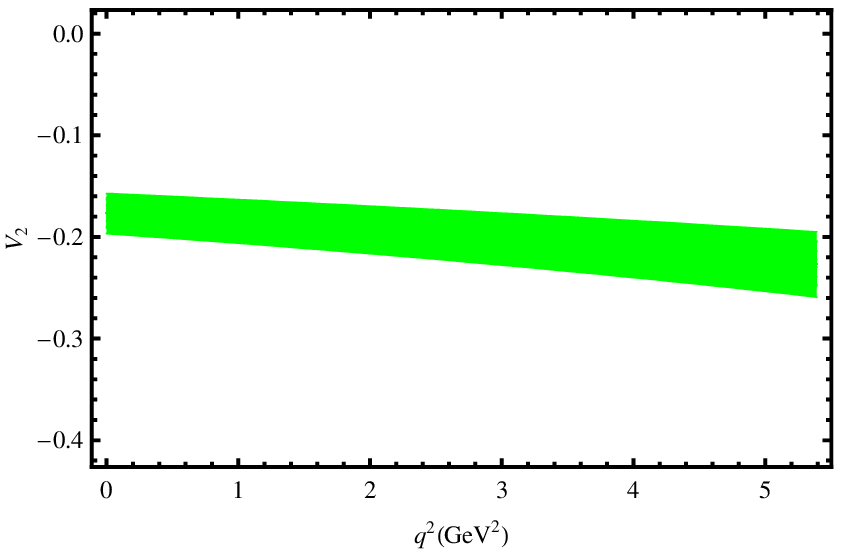}
\caption{The $q^2$-dependence of the $B_s\to D_s(3040)$ form factors
with the quantum numbers $P^{1/2}_1$}\label{fig:q2-Ds(3040)A-1over2}
\end{figure}

\begin{figure}
\includegraphics[scale=0.8]{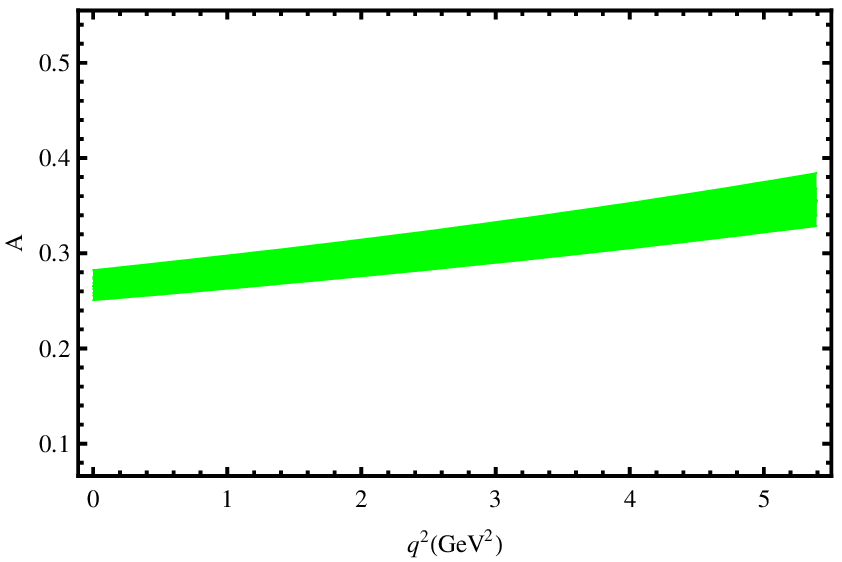}
\includegraphics[scale=0.8]{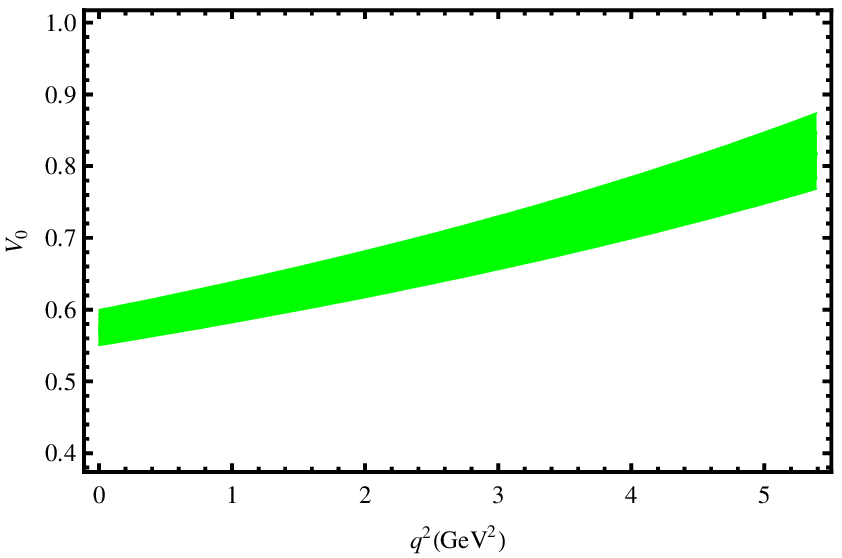}
\includegraphics[scale=0.8]{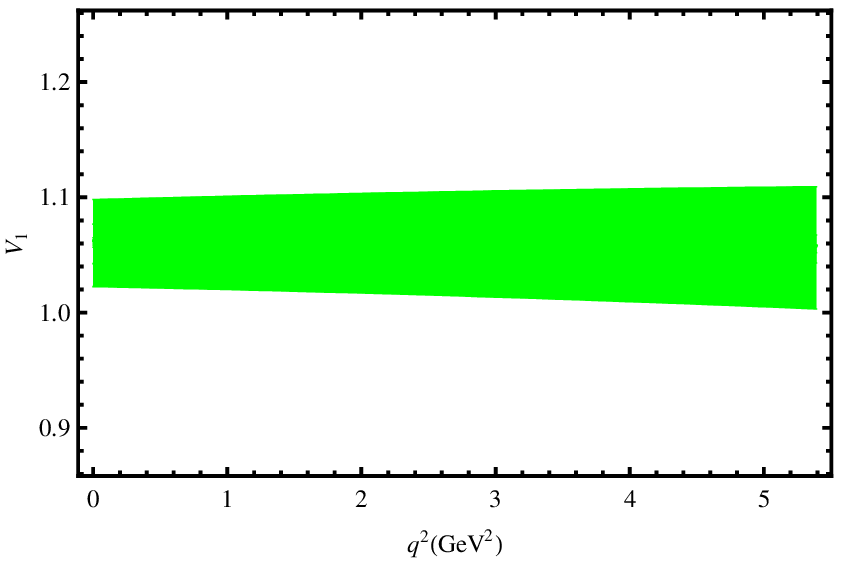}
\includegraphics[scale=0.8]{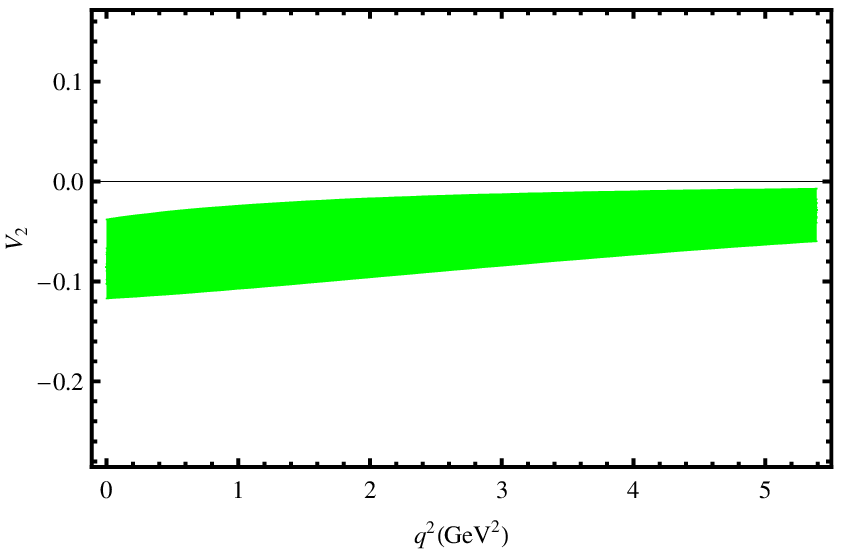}
\caption{The $q^2$-dependence of the $B_s\to D_s(3040)$ form factors
with the quantum numbers $P^{3/2}_1$}\label{fig:q2-Ds(3040)A-3over2}
\end{figure}

Several remarks are given in order. Firstly, the $B_s\to D_s^*$ form
factors are related to the $B\to D^*$ form factors by the flavor
SU(3) symmetry. Compared with the computation in
Ref.~\cite{Cheng:2003sm}
\begin{eqnarray}
 V ^{BD^*}(q^2)=\frac{0.75}{1-1.29 q^2/m_{B}^2+0.45(q^2/m_{B}^2)^2
 },\;\;\;
 A_0^{BD^*}(q^2)=\frac{0.64}{1-1.30 q^2/m_{B}^2+0.31(q^2/m_{B}^2)^2
 },\nonumber\\
 A_1^{BD^*}(q^2)=\frac{0.63}{1-0.65 q^2/m_{B}^2+0.02(q^2/m_{B}^2)^2
 },\;\;
 A_2^{BD^*}(q^2)=\frac{0.61}{1-1.14 q^2/m_{B}^2+0.52(q^2/m_{B}^2)^2
 },
\end{eqnarray}
we can see that our predictions are compatible with their results.
The small differences originate from the SU(3) symmetry breaking
effects in the constituent quark, hadron masses, and decay constants
of the initial and  final mesons.

From the results in table~\ref{tab:form-factors}, we can see that
although most $B_s\to D_{s1}$ form factors are smaller than the
relevant $B_s\to D_s^*$ ones, they are still large enough. It
implies that the $B_s\to D_{s1}$ decay channels are very likely to
be observable. In particular,  the large form factors
$V_1^{B_sD_{s1}(P^{3/2}_1)}$ and $V_1^{B_sD_{s1}(^1P_1)}$ directly
predict the sizable production rates in nonleptonic $B_s$ decays,
which will be shown in the following section.

One important feature of the $B_s\to D_s^*$ form factors is that
they become larger with the increasing $q^2$. However, not all the
$B_s\to D_{s1}$ form factors have this universal behavior. For
instance, the form factor $V_1^{B_sD_{s1}(^3P_1)}$ decreases with
the increasing $q^2$.

Most form factors have the uncertainties of the order $10\%$ but
some of them can reach $20\%$. The uncertainties are denoted by the
bands in Figs.(3-6). The $q^2$-dependent parameters of
$V_2^{B_sD_s(A^{3/2})}$ also suffer from large uncertainties. It is
fortunate that the magnitude of this form factor is small. Thus, it
has little influence on physical processes.

\begin{table}
\caption{$B_s\to D_s^*$ and $B_s\to D_s(3040)$ form factors. Four
sets of results are shown for $D_s(3040)$, with different
assignments of quantum numbers. }\label{tab:form-factors}
\begin{tabular}{|c|c|c|c|c|c}
\hline\hline & $F_i(q^2=0)$ & $F_i(q^2_{\rm max})$  & $a_i$ & $b_i$
\\\hline
 $V^{B_sD^{*}_s}$   & $0.74^{+0.05}_{-0.05}$ & $1.27^{+0.13}_{-0.12}$ &$1.34^{+0.11}_{-0.10}$ &$0.57^{+0.15}_{-0.13}$ \\
 $A_0^{B_sD^{*}_s}$ & $0.63^{+0.04}_{-0.04}$ & $1.05^{+0.06}_{-0.06}$ &$1.30^{+0.09}_{-0.09}$ &$0.54^{+0.14}_{-0.12}$ \\
 $A_1^{B_sD^{*}_s}$ & $0.61^{+0.03}_{-0.03}$ & $0.82^{+0.05}_{-0.05}$ &$0.71^{+0.09}_{-0.08}$ &$0.08^{+0.05}_{-0.04}$ \\
 $A_2^{B_sD^{*}_s}$ & $0.59^{+0.04}_{-0.04}$ & $0.97^{+0.09}_{-0.08}$ &$1.24^{+0.10}_{-0.09}$ &$0.48^{+0.13}_{-0.11}$ \\
 \hline\hline
 $A^{B_sD_{s1}(^3P_1)}$&$0.29^{+0.03}_{-0.03}$ & $0.39^{+0.05}_{-0.05}$ &$1.36^{+0.12}_{-0.12}$ &$0.26^{+0.06}_{-0.04}$ \\
 $V_0^{B_sD_{s1}(^3P_1)}$& $0.27^{+0.02}_{-0.02}$&$0.37^{+0.02}_{-0.02}$&$1.52^{+0.10}_{-0.10}$ &$0.30^{+0.06}_{-0.05}$ \\
 $V_1^{B_sD_{s1}(^3P_1)}$& $0.97^{+0.09}_{-0.10}$&$0.90^{+0.11}_{-0.12}$&$-0.35^{+0.15}_{-0.18}$&$0.33^{+0.14}_{-0.11}$ \\
 $V_2^{B_sD_{s1}(^3P_1)}$& $0.07^{+0.02}_{-0.02}$&$0.07^{+0.02}_{-0.02}$&$-0.25^{+0.39}_{-0.46}$&$0.53^{+0.31}_{-0.23}$ \\
 $A^{B_sD_{s1}(^1P_1)}$ &  $0.12^{+0.01}_{-0.01}$&$0.16^{+0.02}_{-0.02}$&$1.50^{+0.10}_{-0.10}$ &$0.71^{+0.17}_{-0.14}$ \\
 $V_0^{B_sD_{s1}(^1P_1)}$& $0.52^{+0.06}_{-0.06}$&$0.74^{+0.10}_{-0.09}$&$1.73^{+0.11}_{-0.11}$ &$0.48^{+0.09}_{-0.08}$\\
 $V_1^{B_sD_{s1}(^1P_1)}$& $0.62^{+0.06}_{-0.06}$&$0.68^{+0.07}_{-0.07}$&$0.46^{+0.12}_{-0.12}$ &$0.12^{+0.04}_{-0.03}$\\
 $V_2^{B_sD_{s1}(^1P_1)}$& $-0.20^{+0.03}_{-0.04}$&$-0.29^{+0.06}_{-0.06}$&$2.03^{+0.12}_{-0.12}$&$1.94^{+0.32}_{-0.26}$\\
 \hline\hline
 $A^{B_sD_{s1}(P^{1/2}_1)}$&$-0.17^{+0.03}_{-0.02}$&$-0.22^{+0.04}_{-0.04}$&$1.29^{+0.14}_{-0.15}$&$0.14^{+0.06}_{-0.03}$\\
 $V_0^{B_sD_{s1}(P^{1/2}_1)}$&$ 0.08^{+0.05}_{-0.05}$&$0.12^{+0.08}_{-0.09}$&$2.15^{+0.08}_{-0.25}$&$1.8^{+3.7}_{-1.0}$\\
 $V_1^{B_sD_{s1}(P^{1/2}_1)}$&$-0.43^{+0.11}_{-0.10}$&$-0.37^{+0.11}_{-0.11}$&$-0.87^{+0.32}_{-0.43}$&$0.67^{+0.38}_{-0.26}$\\
 $V_2^{B_sD_{s1}(P^{1/2}_1)}$&$-0.18^{+0.03}_{-0.03}$&$-0.23^{+0.05}_{-0.05}$&$1.21^{+0.16}_{-0.19}$&$0.21^{+0.07}_{-0.04}$\\
 $A ^{B_sD_{s1}(P^{3/2}_1)}$ &$0.27^{+0.02}_{-0.02}$ &$0.36^{+0.03}_{-0.03}$ &$1.41^{+0.10}_{-0.10}$&$0.40^{+0.08}_{-0.06}$\\
 $V_0^{B_sD_{s1}(P^{3/2}_1)}$&$0.57^{+0.04}_{-0.04}$ &$0.82^{+0.07}_{-0.07}$ &$1.67^{+0.11}_{-0.11}$&$0.42^{+0.08}_{-0.07}$\\
 $V_1^{B_sD_{s1}(P^{3/2}_1)}$&$ 1.06^{+0.04}_{-0.04}$&$1.06^{+0.05}_{-0.06}$ &$0.01^{+0.07}_{-0.08}$&$0.17^{+0.05}_{-0.04}$\\
 $V_2^{B_sD_{s1}(P^{3/2}_1)}$&$-0.09^{+0.06}_{-0.04}$&$-0.03^{+0.03}_{-0.04}$&$-6^{+5}_{-10}$ & $25^{+17}_{-10}$\\
 \hline\hline
\end{tabular} \end{table}

\section{applications to $B_s$ decays}

In this section, we will use the above determined form factors
to predict  semilepotnic $B_s\to D_s^*(D_s(3040))\ell\bar\nu$ and
nonleptonic $B_s$ decays. For nonleptonic $B_s$ decays, the
factorization assumption will  be used to decompose the decay
amplitudes into the form factors and the decay constant of the
emitted meson. In particular, we will only consider the
color-allowed decay channels where factorization assumptions work
well and the nonfactorizable contributions are typically small.

\subsection{Semileptonic $B_s$ decays}

With the form factors available,
we can investigate the semileptonic $B_s\to D_s^*$ and $B_s\to
D_s(3040)$ decays
\begin{eqnarray}
 \frac{d\Gamma^{M}}{dq^2}&=&\sum_{i=L,\pm}
 \frac{d\Gamma^{M}_{i}}{dq^2},\nonumber\\
 \frac{d\Gamma^{M}_{L, \pm}}{dq^2}&=&
 \frac{ |G_F V_{cb}|^2  \sqrt  {\lambda_M} }{256m_{B_s}^3\pi^3q^2}
\left(1- {\frac{m_\ell^2}{q^2}}\right)^2
 (X^{M}_{L}, X^{M}_{\pm})
\end{eqnarray}
where $M$ denotes $D_s^*$ or $D_s(3040)$,
$\lambda_M=\lambda(m^2_{B_s},m^2_M, q^2)$, and
$\lambda(a^2,b^2,c^2)=(a^2-b^2-c^2)^2-4b^2c^2$. The subscript
$(L,\pm)$ denotes the three polarizations of the vector and
axial-vector $\bar cs$ meson along its momentum direction
$(0,\pm1)$. In terms of the angular distributions, we can study the
forward-backward asymmetries (FBAs) of lepton which are defined as
 \begin{eqnarray}
 \frac{dA^{M}_{FB}}{dq^2} &=& \frac{\int^{1}_{0} dz (d\Gamma^M/dq^2dz) -
\int^{0}_{-1} dz (d\Gamma^M/dq^2dz)}{\int^{1}_{0} dz
(d\Gamma^M/dq^2dz) + \int^{0}_{-1} dz (d\Gamma^M/dq^2dz)}\nonumber
 \end{eqnarray}
where $z\equiv \cos\theta$ and the angle $\theta$ is the polar angle
of the lepton with respect to the moving direction of
$D^{*}_s(D_{s1})$ in the lepton pair rest frame. Explicitly, we have
\begin{eqnarray}
 \frac{dA_{\rm FB}^{D_s^*}}{dq^2}&=&
 \frac{1}{X^{D_s^*}_L + X^{D_s^*}_{+} + X^{D_s^*}_{-}} \left( 2  m_\ell^2
 \sqrt{\lambda_{D_s^*}}h_0^{D_s^*}(q^2) A_0(q^2)  -4q^4\sqrt {\lambda_{D_s^*}} A_1(q^2) V(q^2)
 \right),\nonumber\\
 \frac{dA_{\rm FB}^{D_{s1}}}{dq^2}&=&
 \frac{1}{X^{D_{s1}}_L + X^{D_{s1}}_{+} + X^{D_{s1}}_{-}} \left( 2  m_\ell^2
 \sqrt{\lambda_{D_{s1}}}h_0^{D_{s1}}(q^2) V_0(q^2)  -4q^4\sqrt {\lambda_{D_{s1}}} V_1(q^2) A(q^2)
 \right),\label{eq:AAS-Ds1}
\end{eqnarray}
where
\begin{eqnarray}
X^{D^*_s}_{L} &=& \frac{2}{3} \left[ (2q^2+m_\ell^2)
(h_0^{D_s^*}(q^2))^2 + 3
{\lambda_{D^*_s}} m_\ell^2 A_0^2 (q^2)\right]\,, \nonumber \\
X^{D^*_s}_{\pm} &=& \frac{2q^2}{3}  (2 q^2 + m_\ell^2 )
\left[(m_{B_s}+m_{D^*_s})A_1(q^2) \mp \frac{\sqrt{\lambda_{D^*_s}}
}{m_{B_s}+m_{D^*_s}}V(q^2) \right]^2,\nonumber\\
 h_0^{D_s^*}(q^2)&=& \frac{ 1}{2 m_{D^*_s}
}\left[(m_{B_s}^2-m_{D^*_s}^2-q^2)(m_{B_s}+m_{D^*_s})A_1(q^2)-
\frac{{\lambda_{D^*_s}
}}{m_{B_s}+m_{D^*_s}}A_2(q^2)\right],\nonumber\\
X^{D_{s1}}_{L} &=& \frac{2}{3} \left[ (2q^2+m_\ell^2)
(h_0^{D_{s1}}(q^2))^2 + 3
{\lambda_{D_{s1}}} m_\ell^2 V_0^2(q^2) \right]\,, \nonumber \\
X^{D_{s1}}_{\pm} &=& \frac{2q^2}{3}  (2 q^2 + m_\ell^2 )
\left[(m_{B_s}-m_{D_{s1}})V_1(q^2) \mp \frac{\sqrt{\lambda_{D_{s1}}}
}{m_{B_s}-m_{D_{s1}}}A(q^2) \right]^2,\nonumber\\
 h_0^{D_{s1}}(q^2)&=& \frac{ 1}{2 m_{D_{s1}}
}\left[(m_{B_s}^2-m_{D_{s1}}^2-q^2)(m_{B_s}-m_{D_{s1}})V_1(q^2)-
\frac{{\lambda_{D_{s1}}
}}{m_{B_s}-m_{D_{s1}}}V_2(q^2)\right].\label{eq:longitudinal-Ds1}
\end{eqnarray}
Integrating over the $q^2$, we obtain the partial decay width and
integrated angular asymmetry  for the $B_s\to D_{sJ}$ decay modes:
\begin{eqnarray}
 \Gamma^M=\Gamma_L^M+\Gamma_+^M+\Gamma_-^M,\;\;\;
 A^{M}_{FB}  &=& \frac{1}{\Gamma^M}\int dq^2 \int^{1}_{-1}sign(z) dz (d\Gamma^M/dq^2dz) \nonumber
\end{eqnarray}
with
\begin{eqnarray}
 \Gamma^M_{L,\pm}=\int_{m_l^2}^{(m_{B_s}-m_{M})^2} dq^2
 \frac{d\Gamma_{L,\pm}^M}{dq^2}.
\end{eqnarray}
Since there are three different polarizations,  it is also
meaningful to define the polarization fraction
\begin{eqnarray}
 f_L=\frac{\Gamma_L}{\Gamma_L+\Gamma_++\Gamma_-}.
\end{eqnarray}

\begin{table}
\caption{Branching ratios, polarization fractions and angular
asymmetries for $B_s\to D_s^*$ and $B_s\to D_s(3040)$}\label{tab:semileptonic}
\begin{tabular}{|c|c|c|c|c|c|c}
\hline\hline
 &  $\ell=\mu$  & $\ell=\tau$    \\\hline
 ${\cal B}(\bar B^0_s\to D^{*+}_s \ell\bar\nu_\ell)$ &
 $(5.2 ^{+0.6}_{-0.6})\times 10^{-2}$& $(1.3^{+0.2}_{-0.1})\times 10^{-2}$ \\
 $f_L(\bar B^0_s \to  D^{*+}_s \ell\bar\nu_\ell)$&    $(51.3^{+1.2}_{-1.3})\times 10^{-2}$
 &$(44.5^{+0.8}_{-0.9})\times 10^{-2}$ \\
  $A_{\rm FB}(\bar B^0_s \to  D^{*+}_s \ell\bar\nu_\ell)$    &   $(
  -21.7^{+1.4}_{-1.5})\times 10^{{-2}})$   &    $(-5.6^{+1.4}_{-1.5})\times 10^{{-2}}$
  \\\hline
 ${\cal B}(\bar B^0_s \to  D^+_{s1}(^3P_1) \ell\bar\nu_\ell)$      &   $(
  2.3^{+0.3}_{-0.3})\times 10^{{-3}}$   &    $(6.1^{+1.2}_{-1.2})\times 10^{{-5}}$      \\
 $f_L(\bar B^0_s \to  D^+_{s1}(^3P_1) \ell\bar\nu_\ell)$   &   $(
  43.3^{+3.9}_{-4.2})\times 10^{{-2}}$   &    $(36.0^{+1.9}_{-2.2})\times 10^{{-2}}$        \\
 $A_{\rm FB}(\bar B^0_s \to  D^+_{s1}(^3P_1) \ell\bar\nu_\ell)$     &   $(
  -39.7^{+4.2}_{-4.0})\times 10^{{-2}}$   &    $(-16.6^{+4.0}_{-4.2})\times 10^{{-2}}$     \\
 ${\cal B}(\bar B^0_s \to  D^+_{s1}(^1P_1) \ell\bar\nu_\ell)$    &   $(
  2.9^{+0.6}_{-0.6})\times 10^{{-3}}$   &    $(5.2^{+1.2}_{-1.0})\times 10^{{-5}}$        \\
 $f_L(\bar B^0_s \to  D^+_{s1}(^1P_1) \ell\bar\nu_\ell)$    &   $(
  84.0^{+2.2}_{-2.3})\times 10^{{-2}}$   &    $(65.9^{+3.1}_{-3.0})\times 10^{{-2}}$        \\
 $A_{\rm FB}(\bar B^0_s \to  D^+_{s1}(^1P_1) \ell\bar\nu_\ell)$   &   $(
  -6.8^{+1.4}_{-1.5})\times 10^{{-2}}$   &    $(19.0^{+2.5}_{-2.6})\times 10^{{-2}}$         \\
\hline\hline
 ${\cal B}(\bar B^0_s \to  D^+_{s1}(P^{1/2}_1) \ell\bar\nu_\ell)$  &      $(
  3.5^{+1.1}_{-1.0})\times 10^{{-4}}$   &    $(9.9^{+4.4}_{-3.5})\times 10^{{-6}}$     \\
$f_L(\bar B^0_s \to  D^+_{s1}(P^{1/2}_1) \ell\bar\nu_\ell)$    &
$(
  9.8^{+11.9}_{-3.4})\times 10^{{-2}}$   &    $(18.0^{+2.3}_{-0.2})\times 10^{{-2}}$     \\
$A_{\rm FB}(\bar B^0_s \to  D^+_{s1}(P^{1/2}_1) \ell\bar\nu_\ell)$
& $(
  -65.0^{+13.4}_{-3.2})\times 10^{{-2}}$   &    $(-47.6^{+8.6}_{-2.3})\times 10^{{-2}}$    \\
${\cal B}(\bar B^0_s \to  D^+_{s1}(P^{3/2}_1) \ell\bar\nu_\ell)$     &
$(
  4.0^{+0.4}_{-0.5}\times 10^{{-3}}$   &    $(9.7^{+0.8}_{-0.8})\times 10^{{-5}}$    \\
$f_L(\bar B^0_s \to  D^+_{s1}(P^{3/2}_1) \ell\bar\nu_\ell)$    &   $(
  63.9^{+3.5}_{-4.0})\times 10^{{-2}}$   &    $(49.9^{+1.8}_{-1.6})\times 10^{{-2}}$    \\
$A_{\rm FB}(\bar B^0_s \to  D^+_{s1}(P^{3/2}_1) \ell\bar\nu_\ell)$
& $(
  -22.2^{+1.9}_{-2.1})\times 10^{{-2}}$   &    $(2.0^{+0.7}_{-0.5})\times 10^{{-2}}$    \\
\hline\hline
\end{tabular} \end{table}

Our results for the branching fractions, FBAs and polarizations are
given in Table~\ref{tab:semileptonic}, where $V_{cb}=0.0412$ is
employed~\cite{Amsler:2008zzb}. The uncertainties are from the form
factors. Several remarks are given in order. Our result  for the
${\cal B}(B_s\to D_s^* l\bar\nu)$  is well consistent with  the QCD
sum rules in Ref.~\cite{Blasi:1993fi},
 ${\cal B}(B_s\to D_s^* l\bar\nu)=(5.6\pm0.2)\%$,
but is smaller than the prediction in Ref.~\cite{Zhang:2010ur} by
roughly $40\%$,
 ${\cal B}(B_s\to D_s^* e\bar\nu)=(7.09\pm0.88)\%$.
At present there are no experimental data on the semileptonic $B_s$
decays into $D_s^*$.  Interestingly, our predictions are close to
the $B\to D^* l\bar\nu$ data~\cite{Amsler:2008zzb} which can be
related to the $B_s$ decays by the flavor SU(3) symmetry
\begin{eqnarray}
 {\cal B}(\bar B^0\to D^{*+}l^-\bar\nu) &=& (5.16\pm0.11)\%,\;\;\;
 {\cal B}(\bar B^0\to D^{*+}\tau^-\bar\nu_\tau) = (1.6\pm0.5)\%.
\end{eqnarray}
Since the flavor symmetry breaking
is at a few percent level, our results are also supported by the
nonleptonic $B$ and $B_s$ decay data~\cite{Amsler:2008zzb}
\begin{eqnarray}
 {\cal B}(\bar B^0\to D^+\pi^-)&=& (2.68\pm0.23)\times
 10^{-3}\sim
 {\cal B}(\bar B^0_s\to D^+_s\pi^-)= (3.2\pm0.9)\times
 10^{-3}.
\end{eqnarray}
Furthermore our results show that the polarization fractions and
angular asymmetries are sizable and can be tested at the hadron
collider and the Super B experiment in the near future.



Compared with $B_s\to D_s^* l\bar\nu$, the BRs of the  $B_s\to
D_s(3040) l\bar\nu$ ($l=\mu$) decays are typically smaller by one
order of magnitude. However, if $D_{s}(3040)$ has the quantum number
$P^{1/2}_1$, the BR would be suppressed  by two orders of magnitude
as a consequence of the tiny form factors. The mass of the
$D_{s}(3040)$ is larger than that of $D_s^*$, making the phase space
in $B_s\to D_{s1}\tau\bar\nu_\tau$ much smaller and inducing further
suppressions of the BRs. Nevertheless, the BRs are still large
enough to measure in the future. For example, the LHCb experiment
will produce more than $10^{6}$ $B_s$ mesons per running year,
implying that a plenty of the $D_s(3040)$ events will be generated.

Although the production rates for  $D_s(3040)$ under the two
different descriptions $^3P_1$ and $^1P_1$ are similar, their
polarizations are quite different. For the $^3P_1$ assignment, the
form factor $V_1$ is much larger than the other ones. The
interference between different form factors does not manifest. Since
both longitudinal and transverse polarizations depend on this large
$V_1$, the polarization fraction is close to $50\%$. For $^1P_1$,
the magnitude of $V_2$ is enhanced and it has a different sign with
$V_1$. These two form factors will provide constructive
contributions to the longitudinal polarization as shown in
Eq.~(\ref{eq:longitudinal-Ds1}), and lead to a larger value for
$f_L$.

The angular asymmetries in $B_s\to D_{s1}l\bar\nu$ with
$D_{s1}=(^3P_1,^1P_1)$ are also different. If the lepton is a light
muon, the mass terms in the angular asymmetries as in
Eq.~(\ref{eq:AAS-Ds1}) are negligible. Thus, the form factor product
$V_1(q^2)A(q^2)$ determines the size of the angular asymmetries. For
$^3P_1$, the product is much larger than that of the $^1P_1$
assignment. So is the angular asymmetry, but with a negative sign.
If the lepton is changed to a massive tau, the two terms in
Eq.~(\ref{eq:AAS-Ds1}) give destructive contributions, which will
reduce the modulo of the angular asymmetries  for $^3P_1$ and change
the sign for the $^1P_1$.

The above analysis is also similar for the other
assignment basis. One exceptional case is the polarization fraction for the
$P^{1/2}_1$. All form factors are relatively small, and the
cancelation between $V_1$ and $V_2$ in the longitudinal polarization
results in a small polarization fraction $f_L$.

\subsection{Nonleptonic $B_s$ decays}

After the contraction of the highly-virtual degrees of freedom such
as the $W,Z$ bosons, the effective Hamiltonian governing nonleptonic
$B_s\to D_{sJ}$ decays is given as
\begin{eqnarray}
 H_{\rm eff} &=& \frac{G_F}{\sqrt 2} \sum_{p,D} V_{cb} V_{pD}^*
 (C_1 O_1+C_2O_2)+ h.c.,
\end{eqnarray}
with the four quark operators
\begin{eqnarray}
 O_1=[\bar c^\alpha\gamma^\mu(1-\gamma_5)b^\alpha][\bar
 D^\beta\gamma_\mu(1-\gamma_5)p^\beta],\;\;\
 O_2=[\bar c^\alpha\gamma^\mu(1-\gamma_5)b^\beta][\bar
 D^\beta\gamma_\mu(1-\gamma_5)p^\alpha],
\end{eqnarray}
and $C_1$ and $C_2$ being their Wilson coefficients which arise from
the perturbative coefficients over the $m_b$ scale. We use
$a_1=C_1+C_2/3=1.07$~\cite{Buchalla:1995vs}. In the above equation,
it is labeled $p=u,c$ and $D=d,s$, while $\alpha$ and $\beta$ are
color indices. The CKM matrix elements are employed as
$V_{ud}=0.97418$, $|V_{cd}|=0.23$, $V_{cs}=0.997$ and
$V_{us}=0.2255$~\cite{Amsler:2008zzb}.

Nonleptonic $B_s\to D_{sJ}M$ decays involve one more meson and the
two final state mesons will get entangled with each other. In case
of the $M$ being a light meson, it is feasible to show that the
decay amplitude can be decomposed into two individual parts. This
method is known as factorization~\cite{bsw,Ali:1997nh,9804363}. The
light meson moves very fast in the $B_s$ rest frame and it is made
of collinear quark fields and gluon fields. On the contrary, the
$B_s$ meson and the recoiled $D_{sJ}$ are almost at rest and the
dynamic degree of freedom is the heavy quark, the light antiquark
with quantum fluctuations. The complete set of the degrees of
freedom in these heavy mesons has typically small
momentum~\footnote{For the heavy quark, it means the residual
momentum after separating the large label momentum.} and is
characterized by the soft fields. Interactions between the two
different sectors are forbidden at the leading power and these two
sectors are dynamically separated. The elegant proof rooted in the
quantum field theory  has been given in Ref.~\cite{Bauer:2001cu},
where an effective field theory built in Ref.~\cite{Bauer:2000yr}
has been used.
%
%


Under such an effective theory, the matrix element of the four-quark
operators in two-body $B_s$ decays is expressed in terms of the
$B_s\to D_{sJ}$ form factors, which have been computed in the above,
together with the convolution of a hard kernel $T(u)$ and the
light-cone distribution amplitude $\phi(u)$ of the light meson. The
hard kernel can be computed in the expansion of $\alpha_s$, and
particularly at the leading order $T(u)=1+{\cal O}(\alpha_s)$. In
this case, the analysis based on the effective theory recovers the
naive factorization in which the decay amplitudes for $\overline
B_s\to D_{s}^{*+}\pi^-$ and $\overline B_s\to D_{s}^{*+}\rho^-$ are
given as
\begin{eqnarray}
 {\cal A}( \overline B_s^0\to D_s^{*+} \pi^-)&=& \frac{G_F}{\sqrt 2}
 V_{cb}V_{ud}^* a_1
 f_{\pi} A_0(m_{\pi}^2)
 \sqrt {\lambda(m_{B_s}^2, m_{D_s^*}^2,m_{\pi}^2)},\nonumber\\
 {\cal A}_L(\overline B_s^0\to D_s^{*+}\rho^-)&=& \frac{-i G_F}{\sqrt 2}
 V_{cb}V_{ud}^* a_1   f_{\rho}h_0^{D_s^*}(m_\rho^2),\nonumber\\
 {\cal A}_N(\overline B_s^0\to D_s^{*+} \rho^-)&=&
 \frac{-i G_F}{\sqrt 2}
 V_{cb}V_{ud}^* a_1  f_{\rho}(m_{B_s}+m_{D_s^*})   m_{\rho} A_1(m_{\rho}^2),\nonumber\\
 {\cal A}_T(\overline B_s^0\to {D_s^{*+}} \rho^-)&=&
 \frac{-i G_F}{\sqrt 2}
 V_{cb}V_{ud}^* a_1   f_{\rho}  \frac{\sqrt {\lambda(m_{B_s}^2, m_{\rho}^2,m_{D_s^*}^2)}}{(m_{B_s}+m_{D_s^*})}
 m_{\rho}V(m_{\rho}^2).
\end{eqnarray}
In the above, the amplitude of $\overline B_s\to
D_s^{*+}(P_{D_s^*})\rho^-(P_\rho)$ has been decomposed according to
the Lorentz structures
\begin{eqnarray}
 {\cal A}&=& {\cal A}_{L} +\epsilon_{D_s^*}^*(T)\cdot \epsilon_{\rho}^*(T)
 {\cal A}_N + i {\cal
 A}_T\epsilon_{\alpha\beta\gamma\rho}\epsilon^{*\alpha}_{D_s^*}\epsilon^{*\beta}_{\rho}
 \frac{2P_{D_s^*}^{\gamma}P_{\rho}^\rho}
 {\sqrt {\lambda(m_{B_s}^2, m_{D_s^*}^2,m_{\rho}^2)}}.
\end{eqnarray}
We will also consider the decay channels in which the light meson
$(\pi^-,\rho^-)$ is replaced by $K^-, D^-,D_s^-$ or $K^{*-},
D^{*-},D_s^{*-}$. The partial decay width of $\overline B_s\to D_s^*
P$, where P denotes a pseudoscalar meson, is given by
\begin{eqnarray}
 \Gamma(\overline B_s\to D_s^* P)&=& \frac{|\vec p|}{8\pi m_{B_s}^2} \left|{\cal
 A}(\overline B_s\to D_s^* P)\right|^2,
\end{eqnarray}
with $|\vec p|$ being the three-momentum of the $D_s^*$ in the $B_s$
meson rest frame.  For $\overline B_s\to D_s^* V$, the partial decay
width is the summation of the three polarizations
\begin{eqnarray}
 \Gamma(\overline B_s\to D_s^*V)&=& \frac{|\vec p|}{8\pi m_{B_s}^2} \left(\left|{\cal
 A}_0(\overline B_s\to D_s^*V)\right|^2+2\left|{\cal
 A}_{N}(\overline B_s\to D_s^*V)\right|^2+2\left|{\cal
 A}_{T}(\overline B_s\to D_s^*V)\right|^2\right).
\end{eqnarray}
In the decay amplitudes for the channels involving $D_{s1}$, the
form factors are replaced correspondingly and
$m_{B_s}+m_{D_s^*}$ is replaced by $m_{B_s}-m_{D_{s1}}$.

Before proceeding, it is worth mentioning that the proof of factorization of
$B_s$ decays into two charmed mesons such as $B_s\to D_s^* D_s^*$ is
absent in the literature. Nevertheless, since only the color-allowed
tree diagrams are considered in this work, the nonfactorizable
diagrams are typically small and the factorization usually works
very well~\cite{bsw,Ali:1997nh,9804363}. For instance, our
predictions of the branching fractions of these channels are well
consistent with the experimental results. This verifies the
applicability of the factorization approach.

\begin{table}
\caption{Branching ratios of nonleptonic $B_s$ decays into $D_s^*$
and a comparison with theoretical
results\cite{Deandrea:1993ma,Li:2008ts,Li:2009xf} and experimental
data }\label{tab:BR-nonleptonic-comparison}
\begin{tabular}{|c|c|c|c|c|c|c}
\hline\hline  & Factorization~\cite{Deandrea:1993ma}&
 perturbative QCD~\cite{Li:2008ts,Li:2009xf} & This work & Exp.
\\\hline
 $\bar B^0_s \to  D^{*+}_s\pi^-$ &$2.8\times 10^{-3}$  & $(2.42^{+1.37}_{-1.06})\times 10^{-3}$
 &   $(3.5^{+0.4}_{-0.4})\times 10^{{-3}}$  & $(2.4^{+0.7}_{-0.6})\times 10^{-3}$\\
 $\bar B^0_s \to  D^{*+}_sK^-$  &$2.1\times 10^{-4}$ & $(1.65^{+1.06}_{-0.82})\times 10^{-4}$ &    $(
  2.8^{+0.3}_{-0.3})\times 10^{{-4}}$   \\
 $\bar B^0_s \to  D^{*+}_sD^-$ &$3.3\times 10^{-4}$  &$(2.7^{+1.9}_{-1.4})\times 10^{-4}$  &       $(3.7^{+0.4}_{-0.4})\times 10^{{-4}}$ \\
 $\bar B^0_s \to  D^{*+}_sD_s^-$&$7.4\times 10^{-3}$ &$(7.0^{+4.8}_{-3.8})\times 10^{-3}$    &$(9.2^{+1.1}_{-1.1})\times
 10^{{-3}}$\\
 $\bar B^0_s \to  D^{*+}_s\rho^- $  &$8.9\times 10^{-3}$ &$(5.69^{+3.59}_{-2.84})\times 10^{-3}$
 &$(11.8^{+3.3}_{-3.1})\times 10^{-3}$
  &   $(11.3^{+1.4}_{-1.3})\times 10^{{-3}}$  \\
 $\bar B^0_s \to  D^{*+}_s K^{*-}$ &$4.8\times 10^{-4}$&$(3.47^{+2.24}_{-1.72})\times 10^{-4}$  &
 $(5.5^{+0.6}_{-0.6})\times 10^{{-4}}$   \\
 $\bar B^0_s \to  D^{*+}_s D^{*-}$ &$1.0\times 10^{-3}$&$(3.9^{+2.9}_{-2.3})\times 10^{-4}$    &       $(8.6^{+1.0}_{-0.9})\times 10^{{-4}}$    \\
 $\bar B^0_s \to  D^{*+}_s D_{s}^{*-}$ &$2.9\times 10^{-2}$   &$(9.9^{+7.7}_{-6.2})\times 10^{-3}$    &   $(
  23.6^{+4.0}_{-3.8})\times 10^{{-3}}$ & $(3.1^{+1.4}_{-1.3})\%$\\
\hline\hline
\end{tabular} \end{table}

\begin{table}
\caption{Branching ratios of nonleptonic $B_s$ decays into
$D_{s}(3040)$ with different assignments}\label{tab:BR-nonleptonic}
\begin{tabular}{|c|c|c|c|c|c|c}
\hline\hline & $\pi^-$      & $K^-$  &$D^-$ &$D^-_s$   \\\hline%
 $\bar B^0_s \to  D^+_{s1}(^3P_1)$  &$(3.2^{+0.5}_{-0.5})\times10^{{-4}}$    &   $(
  2.5^{+0.4}_{-0.4})\times 10^{{-5}}$   &    $(1.4^{+0.2}_{-0.2})\times 10^{{-5}}$    &   $(
  2.8^{+0.4}_{-0.4})\times 10^{{-4}}$\\
 $\bar B^0_s \to  D^+_{s1}(^1P_1)$  &   $(1.2^{+0.3}_{-0.3})\times 10^{{-3}}$    &   $(
  9.3^{+2.2}_{-2.0})\times 10^{{-5}}$   &    $(5.7^{+1.5}_{-1.3})\times 10^{{-5}}$    &   $(
  1.1^{+0.3}_{-0.3})\times 10^{{-3}}$\\
\hline\hline
 $\bar B^0_s \to  D^+_{s1}(P^{1/2}_1)$  &   $(3.0^{+4.5}_{-2.9})\times10^{{-5}}$
 &   $(  2.3^{+3.5}_{-2.3})\times 10^{{-6}}$   &    $(1.5^{+2.2}_{-1.5})\times 10^{{-6}}$
    &   $(3.0^{+4.4}_{-3.1})\times 10^{{-5}}$\\
 $\bar B^0_s \to  D^+_{s1}(P^{3/2}_1)$  &   $(1.5^{+0.2}_{-0.2})\times10^{{-3}}$ &  $(1.2^{+0.2}_{-0.1})\times 10^{{-4}}$
      &    $(6.9^{+1.1}_{-1.0})\times 10^{{-5}}$       &   $(  1.4^{+0.2}_{-0.2})\times
      10^{{-3}}$\\
\hline\hline  & $\rho^-$      & $K^{*-}$  &$D^{*-}$ &$D^{*-}_s$
\\\hline
 $\bar B^0_s \to  D^+_{s1}(^3P_1)$  &   $(1.3^{+0.2}_{-0.2})\times10^{{-3}}$
   &   $(  6.4^{+0.7}_{-0.8})\times 10^{{-5}}$   &
 $(6.6^{+1.2}_{-1.2})\times 10^{{-5}}$    &   $(
  1.4^{+0.3}_{-0.3})\times 10^{{-3}}$\\
 $\bar B^0_s \to  D^+_{s1}(^1P_1)$  &   $(3.0^{+0.7}_{-0.6})\times 10^{{-3}}$
 &   $( 1.4^{+0.3}_{-0.3})\times 10^{{-4}}$   &    $(4.5^{+0.9}_{-0.8})\times 10^{{-5}}$
   &   $(9.0^{+2.0}_{-1.8})\times 10^{{-4}}$\\
\hline\hline
 $\bar B^0_s \to  D^+_{s1}(P^{1/2}_1)$  &  $(1.6^{+1.0}_{-0.7})\times10^{{-4}}$  &   $(8.5^{+4.4}_{-3.0})\times 10^{{-6}}$
   &  $(1.1^{+0.5}_{-0.4})\times 10^{{-5}}$    &   $(  2.4^{+1.1}_{-1.0})\times
   10^{{-4}}$\\
 $\bar B^0_s \to  D^+_{s1}(P^{3/2}_1)$  &   $(3.3^{+0.6}_{-0.7})\times10^{{-3}}$  & $(1.5^{+0.3}_{-0.3})\times 10^{{-4}}$
   &    $(8.9^{+0.7}_{-0.7})\times 10^{{-5}}$&   $(  1.9^{+0.2}_{-0.2})\times 10^{{-3}}$\\
\hline\hline
\end{tabular} \end{table}


Using the above parameters, we directly obtain the branching
fractions which are collected in
Table~\ref{tab:BR-nonleptonic-comparison} and
Table~\ref{tab:BR-nonleptonic}. For comparison, we also collect two sets of
different theoretical predictions on the $B_s\to D_s^*$
decays~\cite{Deandrea:1993ma,Li:2008ts,Li:2009xf} and the relevant
experimental data~\cite{Louvot:2010rd,Kinoshita:2010kj}. From
Table~\ref{tab:BR-nonleptonic-comparison} we can see that our
predictions on $B_s\to D_s^*$ decay channels are consistent with the
experimental data within uncertainties. This may also imply that our
results for $B_s\to D_s(3040)$ are reliable. Results in the
factorization method~\cite{Deandrea:1993ma} are also close to our
results but the perturbative QCD
predictions~\cite{Li:2008ts,Li:2009xf} are typically smaller. For
instance, the central value of the perturbative QCD result of $\bar
B^0_s \to D^{*+}_s D_{s}^{*-}$ is smaller than our result by a
factor of 2.5.

Our uncertainties are from the form factors, i.e., from the
quark masses and the shape parameters. Uncertainties in the
perturbative QCD approach  shown in
Table~\ref{tab:BR-nonleptonic-comparison} are from three sets of
input parameters: (1) decay constants of $f_{B_s}=(0.24\pm0.03)$ GeV
(in Ref.\cite{Li:2008ts}) or $f_{B_s}=(0.23\pm0.03)$ GeV (in
Ref.\cite{Li:2009xf}), and the shape parameters of the $B_s$ wave
functions $\omega_b=(0.50\pm0.05)$ GeV; (2) the factorization scale
(from 0.75t to 1.25t not changing the transverse part) and the hadronic scale   $\Lambda_{\rm
QCD}=(0.25\pm0.05)$ GeV; (3) the CKM matrix elements $V_{cs}$ and
$V_{ud}$. Among them, the hadronic inputs  are found to give the
largest uncertainties, while the ones from CKM are the smallest. For
instance, the branching fraction of $\bar B_s^0\to D_s^{*+}\pi^-$
with these three kinds of errors is given as
\begin{eqnarray}
 {\cal B}(\bar B_s^0\to D_s^{*+}\pi^-)=(2.42^{+1.12+0.78+0.07}_{-0.72-0.77-0.07})\times 10^{-3}.
\end{eqnarray}

Uncertainties in the PQCD approach are larger than our results for several reasons.
\begin{itemize}
\item Compared with the $f_{B_s}$ adopted in this work,
the uncertainty in their computations is larger by a factor of 2.

\item In the PQCD approach, the $B_s$ wave function
has such a property that the decay constant can be factorized  out
and then $f_{B_s}$ is  only the normalization constant of the wave
functions. In the PQCD approach, the authors have independently
estimated these two uncertainties and added them together as their
final results. On the contrary, in the covariant LFQM the decay
constant can not be factorized, but has been expressed as a
convolution form as shown in Eq.~\eqref{eq:Vdecayconstantusual}. In
this case, only the decay constant is the origin
for the uncertainties, which will be smaller.

\item The uncertainties caused by factorization scales and
hadronic scales in the PQCD approach characterize  higher order QCD
corrections and can be viewed as the model-dependent errors (or
systematic errors). They are smaller than but still in similar
magnitude with  the ones from the $B_s$ wave functions. One
particular type of model-dependent uncertainties in this quark model
concerns the zero-mode contributions. A direct evaluation of the
form factors in the light-front quark model suffers from the
non-covariance problem. It is resolved with the inclusion of the
zero-mode contributions, i.e. the Z-graph as in the conventional
LFQM, and the direct treatment of the spurious
terms with the replacement rules as in the covariant LFQM. Different results can be produced. However, at present
such computation in the conventional light-front quark model is not
available. Therefore, it is not possible to estimate the systematic
errors.
\end{itemize}

In $B_s\to D_{s1}$ decays with a pseudoscalar meson emitted, the
decay amplitudes only involve the form factor $V_0$. The large form
factors for the $P^{3/2}_1$ and $^1P_1$ result in large BRs.
Similarly, the large uncertainty for $V_0(B_s\to D_{s1}(P^{1/2}_1))$
also gives large uncertainties to the BRs.

Channels with the emissions of $K^-$ and $D^-$ are suppressed by
roughly one order of magnitude compared with the case emitting a
pion and $D_s$ meson. This is due to the hierarchy in CKM matrix
elements: $|V_{us}|\sim |V_{cd}|\sim 0.22\ll |V_{ud}|\sim
|V_{cs}|\sim 1$.

Under different quantum number assignments for $D_{s}(3040)$, the
branching fractions $B_s\to D_s(3040)\rho$ and $B_s\to
D_s(3040)D_s^*$ can reach the order of $10^{-3}$. But the modes with
the emissions of $K^*$ and $D^*$ are  smaller by 1 to 2 orders.
Nevertheless, all of these channels have a promising prospect on the
LHCb experiment with sufficient statistics for the $B_s$ production.

\section{Conclusions}




Using the covariant light-front quark model, we have studied the
$B_s\to D_{s}(3040)$ form factors, including the $B_s\to D_s^*$ form
factors as a byproduct. The form factors are computed in the
space-like region and extrapolated to the physical region through a
three-parameter form.    Those form factors are used to predict the
semileptonic and nonleptonic $B_s$ decays. In particular, the
predictions on branching fractions, longitudinal polarization
fractions and the angular asymmetry, are provided. Our results for
$B_s\to D_s^*$ decays are consistent with the experimental data for
the $B_s$ and $B$ decays which are related by the flavor symmetry.
For $B_s\to D_s(3040)$, we find that the branching fractions are
large enough to be observed on the LHCb experiment. For instance,
the BRs of semileptonic $B_s\to D_s(3040)l\bar\nu$, nonleptonic
$B_s\to D_{s1}\rho$, and $B_s\to D_{s1}D_s^*$ can reach the order of
$10^{-3}$. In spite of similar production rates, the different
assignments for the $D_s(3040)$ can be distinguished by the
polarization fractions and angular asymmetries. We expect more
experimental results from the $B$ factories and LHC etc in the near
future. They would be helpful to provide deeper insights into the
$\bar cs$ spectroscopy.

\section*{Acknowledgements}

This work is supported in part by the National Natural Science
Foundation of China under Grant Nos. 10775089, 10805037 and
10947007. W. Wang thanks P. Colangelo for useful discussions and
acknowledges Qufu Normal University for the hospitality during his
visit. We would like to acknowledge Q. Zhao for carefully reading
the manuscript and useful suggestions.

\appendix

\section{Some specific rules under the $p^-$ integration}\label{sec:rules}

When performing the $p^-$ integration one needs to includes the
zero-mode contribution to solve the noncovariance problem. This
amounts to performing the integration in a proper way in the
covariant LFQM~\cite{Jaus:1999zv,Cheng:2003sm}. In particular for
$p_{1\mu}^\prime$ and $\hat p_{1\mu}^\prime p_{1\nu}^\prime $ we
have
 \begin{eqnarray}
\hat p^\prime_{1\mu}
       &\doteq& P_\mu A_1^{(1)}+q_\mu A_2^{(1)},
\hat N_2
       \to Z_2,
  \nonumber\\
\hat p^\prime_{1\mu} \hat p^\prime_{1\nu}
       &\doteq& g_{\mu\nu} A_1^{(2)}+P_\mu P_\nu A_2^{(2)}+(P_\mu
                q_\nu+ q_\mu P_\nu) A^{(2)}_3+q_\mu q_\nu A^{(2)}_4,%
 \label{eq:p1B}
 \end{eqnarray}
with the symbol $\doteq$ denoting that these equations are valid
after integration.  $A^{(i)}_j$ are functions of $x_{1,2}$,
$p^{\prime2}_\bot$, $p^\prime_\bot\cdot q_\bot$ and $q^2$
 \begin{eqnarray}
 Z_2&=&\hat N_1^\prime+m_1^{\prime2}-m_2^2+(1-2x_1)M^{\prime2}
 +(q^2+q\cdot P)\frac{p^\prime_\bot\cdot q_\bot}{q^2},  \nonumber\\
  A^{(1)}_1&=&\frac{x_1}{2},
 \quad%
 A^{(1)}_2=A^{(1)}_1-\frac{p^\prime_\bot\cdot q_\bot}{q^2},
 A^{(2)}_1=-p^{\prime2}_\bot-\frac{(p^\prime_\bot\cdot q_\bot)^2}{q^2},
 \quad%
 \nonumber\\
 A^{(2)}_3&=&A^{(1)}_1 A^{(1)}_2,
 A^{(2)}_4=\big(A^{(1)}_2\big)^2-\frac{1}{q^2}A^{(2)}_1.
 \quad%
 \label{eq:rule}
 \end{eqnarray}


\begin{thebibliography}{99}

\bibitem{Aubert:2003fg}
  B.~Aubert {\it et al.}  [BABAR Collaboration],
  Phys.\ Rev.\ Lett.\  {\bf 90}, 242001 (2003)
  [arXiv:hep-ex/0304021].

\bibitem{Besson:2003cp}
  D.~Besson {\it et al.}  [CLEO Collaboration],
  Phys.\ Rev.\  D {\bf 68}, 032002 (2003)
  [Erratum-ibid.\  D {\bf 75}, 119908 (2007)]
  [arXiv:hep-ex/0305100].




\bibitem{DeFazio:2009xd}
 F.~De Fazio,
  arXiv:0910.0412 [hep-ph].


\bibitem{Evdokimov:2004iy}
  A.~V.~Evdokimov {\it et al.}  [SELEX Collaboration],
  Phys.\ Rev.\ Lett.\  {\bf 93}, 242001 (2004)
  [arXiv:hep-ex/0406045].

\bibitem{Abe:2006xm}
  K.~Abe {\it et al.}  [Belle Collaboration],
  arXiv:hep-ex/0608031.

\bibitem{Aubert:2009di}
  B.~Aubert {\it et al.}  [BABAR Collaboration],
  Phys.\ Rev.\  D {\bf 80}, 092003 (2009)
  [arXiv:0908.0806 [hep-ex]].



\bibitem{Aubert:2006mh}
  B.~Aubert {\it et al.}  [BABAR Collaboration],
  Phys.\ Rev.\ Lett.\  {\bf 97}, 222001 (2006)
  [arXiv:hep-ex/0607082].


\bibitem{Colangelo:2006rq}
  P.~Colangelo, F.~De Fazio and S.~Nicotri,
  Phys.\ Lett.\  B {\bf 642}, 48 (2006)
  [arXiv:hep-ph/0607245].


\bibitem{Zhang:2006yj}
  B.~Zhang, X.~Liu, W.~Z.~Deng and S.~L.~Zhu,
  Eur.\ Phys.\ J.\  C {\bf 50}, 617 (2007)
  [arXiv:hep-ph/0609013].

\bibitem{Chen:2009zt}
  B.~Chen, D.~X.~Wang and A.~Zhang,
  Phys.\ Rev.\  D {\bf 80}, 071502 (2009)
  [arXiv:0908.3261 [hep-ph]].

\bibitem{vanBeveren:2009jq}
  E.~van Beveren and G.~Rupp,
  Phys.\ Rev.\  D {\bf 81}, 118101 (2010)
  [arXiv:0908.1142 [hep-ph]].


\bibitem{Di Pierro:2001uu}
  M.~Di Pierro and E.~Eichten,
  Phys.\ Rev.\  D {\bf 64}, 114004 (2001)
  [arXiv:hep-ph/0104208].



\bibitem{Matsuki:2006rz}
  T.~Matsuki, T.~Morii and K.~Sudoh,
  Eur.\ Phys.\ J.\  A {\bf 31}, 701 (2007)
  [arXiv:hep-ph/0610186].
\bibitem{Ebert:2009ua}
  D.~Ebert, R.~N.~Faustov and V.~O.~Galkin,
  arXiv:0910.5612 [hep-ph].

\bibitem{Close:2006gr}
  F.~E.~Close, C.~E.~Thomas, O.~Lakhina and E.~S.~Swanson,
  Phys.\ Lett.\  B {\bf 647}, 159 (2007)
  [arXiv:hep-ph/0608139].


\bibitem{Wang:2007av}
  G.~L.~Wang,
  Phys.\ Lett.\  B {\bf 650}, 15 (2007)
  [arXiv:0705.2621 [hep-ph]].


\bibitem{Colangelo:2010te}
  P.~Colangelo and F.~De Fazio,
  Phys.\ Rev.\  D {\bf 81}, 094001 (2010)
  [arXiv:1001.1089 [hep-ph]].



\bibitem{Sun:2009tg}
  Z.~F.~Sun and X.~Liu,
  Phys.\ Rev.\  D {\bf 80}, 074037 (2009)
  [arXiv:0909.1658 [hep-ph]].

\bibitem{Zhong:2009sk}
  X.~H.~Zhong and Q.~Zhao,
  Phys.\ Rev.\  D {\bf 81}, 014031 (2010)
  [arXiv:0911.1856 [hep-ph]].




\bibitem{Buchalla:2008jp}
  M.~Artuso {\it et al.},
  Eur.\ Phys.\ J.\  C {\bf 57}, 309 (2008)
  [arXiv:0801.1833 [hep-ph]]; 
%
  N.~Harnew,
  Phys.\ Atom.\ Nucl.\  {\bf 71}, 588 (2008).

\bibitem{Aushev:2010bq}
 For example, see T.~Aushev {\it et al.},
  arXiv:1002.5012 [hep-ex].
The number of $B_s^0$ mesons is estimated to be $\sim5.9\times 10^8$
in the dataset of $L_{\rm int} = 5 ab^{-1}$ taken at the
$\Upsilon(5S)$.


\bibitem{Huang:2004et}
  M.~Q.~Huang,
  Phys.\ Rev.\  D {\bf 69}, 114015 (2004)
  [arXiv:hep-ph/0404032].


\bibitem{Zhao:2006at}
  S.~M.~Zhao, X.~Liu and S.~J.~Li,
  Eur.\ Phys.\ J.\  C {\bf 51}, 601 (2007)
  [arXiv:hep-ph/0612008].
\bibitem{Aliev:2006qy}
  T.~M.~Aliev and M.~Savci,
  Phys.\ Rev.\  D {\bf 73}, 114010 (2006)
  [arXiv:hep-ph/0604002].

\bibitem{Aliev:2006gk}
  T.~M.~Aliev, K.~Azizi and A.~Ozpineci,
  Eur.\ Phys.\ J.\  C {\bf 51}, 593 (2007)
  [arXiv:hep-ph/0608264].

\bibitem{Li:2009wq}
  R.~H.~Li, C.~D.~Lu and Y.~M.~Wang,
  Phys.\ Rev.\  D {\bf 80}, 014005 (2009)
  [arXiv:0905.3259 [hep-ph]].


\bibitem{Jaus:1999zv}
  W.~Jaus,
  Phys.\ Rev.\  D {\bf 60}, 054026 (1999).


\bibitem{Cheng:2003sm}
  H.~Y.~Cheng, C.~K.~Chua and C.~W.~Hwang,
  Phys.\ Rev.\  D {\bf 69}, 074025 (2004).



\bibitem{Cheng:2004yj}
  H.~Y.~Cheng and C.~K.~Chua,
  Phys.\ Rev.\  D {\bf 69}, 094007 (2004)
  [arXiv:hep-ph/0401141].

\bibitem{Ke:2009ed}
  H.~W.~Ke, X.~Q.~Li and Z.~T.~Wei,
  Phys.\ Rev.\  D {\bf 80}, 074030 (2009)
  [arXiv:0907.5465 [hep-ph]].

\bibitem{Ke:2009mn}
  H.~W.~Ke, X.~Q.~Li and Z.~T.~Wei,
  arXiv:0912.4094 [hep-ph].

\bibitem{Cheng:2009ms}
  H.~Y.~Cheng and C.~K.~Chua,
  Phys.\ Rev.\  D {\bf 81}, 114006 (2010)
  [arXiv:0909.4627 [hep-ph]].


\bibitem{Wang:2007sxa}
  C.~D.~Lu, W.~Wang and Z.~T.~Wei,
  Phys.\ Rev.\  D {\bf 76}, 014013 (2007)
  [arXiv:hep-ph/0701265];
  W.~Wang, Y.~L.~Shen and C.~D.~Lu,
  Eur.\ Phys.\ J.\  C {\bf 51}, 841 (2007)
  [arXiv:0704.2493 [hep-ph]];
  W.~Wang and Y.~L.~Shen,
  Phys.\ Rev.\  D {\bf 78}, 054002 (2008); 
  Y.~L.~Shen and Y.~M.~Wang,
  Phys.\ Rev.\  D {\bf 78}, 074012 (2008);
  W.~Wang, Y.~L.~Shen and C.~D.~Lu,
  Phys.\ Rev.\  D {\bf 79}, 054012 (2009)
  [arXiv:0811.3748 [hep-ph]];
  X.~X.~Wang, W.~Wang and C.~D.~Lu,
  Phys.\ Rev.\  D {\bf 79}, 114018 (2009)
  [arXiv:0901.1934 [hep-ph]];
  C.~H.~Chen, Y.~L.~Shen and W.~Wang,
  Phys.\ Lett.\  B {\bf 686}, 118 (2010)
  [arXiv:0911.2875 [hep-ph]];
  W.~Wang,
  arXiv:1002.3579 [hep-ph].







\bibitem{Buchalla:1995vs}
  For a review, see G.~Buchalla, A.~J.~Buras and M.~E.~Lautenbacher,
  Rev.\ Mod.\ Phys.\  {\bf 68}, 1125 (1996)
  [arXiv:hep-ph/9512380].


\bibitem{Jaus:1989au}
  W.~Jaus,
  Phys.\ Rev.\  D {\bf 41}, 3394 (1990).

\bibitem{Jaus:1991cy}
  W.~Jaus,
  Phys.\ Rev.\  D {\bf 44}, 2851 (1991).

\bibitem{Cheng:1996if}
  H.~Y.~Cheng, C.~Y.~Cheung and C.~W.~Hwang,
  Phys.\ Rev.\  D {\bf 55}, 1559 (1997)
  [arXiv:hep-ph/9607332].

\bibitem{Choi:2001hg}
  H.~M.~Choi, C.~R.~Ji and L.~S.~Kisslinger,
  Phys.\ Rev.\  D {\bf 65}, 074032 (2002)
  [arXiv:hep-ph/0110222].



\bibitem{Amsler:2008zzb}
  C.~Amsler {\it et al.}  [Particle Data Group],
  Phys.\ Lett.\  B {\bf 667}, 1 (2008);
 K.~Nakamura  {\it et al.}  [Particle Data Group],
  J.\ Phys.\  G {\bf 37}, 075021 (2010).


\bibitem{Choi:1999vu}
  H.~M.~Choi,
  arXiv:hep-ph/9911271.


\bibitem{Gamiz:2009ku}
  E.~Gamiz, C.~T.~H.~Davies, G.~P.~Lepage, J.~Shigemitsu and M.~Wingate
                  [HPQCD Collaboration],
  Phys.\ Rev.\  D {\bf 80}, 014503 (2009)
  [arXiv:0902.1815 [hep-lat]].


\bibitem{HFAG}
  Heavy-Flavor-Averaging-Group,
  http://www.slac.stanford.edu/xorg/hfag/charm/PIC09/f\_ds/results.html.



\bibitem{Faessler:2007cu}
  A.~Faessler, T.~Gutsche, S.~Kovalenko and V.~E.~Lyubovitskij,
  Phys.\ Rev.\  D {\bf 76}, 014003 (2007)
  [arXiv:0705.0892 [hep-ph]].


%



\bibitem{Louvot:2010rd}
  R.~Louvot {\it et al.}  [Belle Collaboration],
  Phys.\ Rev.\ Lett.\  {\bf 104}, 231801 (2010)
  [arXiv:1003.5312 [hep-ex]].



\bibitem{Kinoshita:2010kj}
  K.~Kinoshita,
  arXiv:1005.3893 [hep-ex];
  S.~Esen {\it et al.} [Belle Collaboration],
  arXiv:1005.5177 [hep-ex].

%



\bibitem{bsw}M. Wirbel, B. Stech, M. Bauer, Z. Phys. C{\bf29}, 637 (1985);
 M. Bauer, B. Stech, M. Wirbel, Z. Phys. C{\bf34}, 103 (1987).

\bibitem{Ali:1997nh}
  A.~Ali and C.~Greub,
  Phys.\ Rev.\ D{\bf 57}, 2996 (1998)
  [hep-ph/9707251];
G.~Kramer, W. F.~Palmer and H.~Simma, Nucl.\ Phys.\ B{\bf 428}, 77
(1994); Z.\ Phys.\ C{\bf 66}, 429 (1995).


 \bibitem{9804363} A. Ali, G. Kramer, and C. -D. L\"{u},
          Phys. Rev. D{\bf 58}, 094009 (1998) [hep-ph/9804363];
          Phys. Rev. D{\bf 59}, 014005 (1999) [hep-ph/9805403];
          Y. H. Chen, H. Y. Cheng, B. Tseng, and K. C. Yang,
          Phys. Rev. D{\bf 60}, 094014 (1999) [hep-ph/9903453].

\bibitem{Bauer:2001cu}
  C.~W.~Bauer, D.~Pirjol and I.~W.~Stewart,
  Phys.\ Rev.\ Lett.\  {\bf 87}, 201806 (2001)
  [arXiv:hep-ph/0107002].

\bibitem{Bauer:2000yr}
  C.~W.~Bauer, S.~Fleming, D.~Pirjol and I.~W.~Stewart,
  Phys.\ Rev.\  D {\bf 63}, 114020 (2001)
  [arXiv:hep-ph/0011336].


\bibitem{Blasi:1993fi}
  P.~Blasi, P.~Colangelo, G.~Nardulli and N.~Paver,
  Phys.\ Rev.\  D {\bf 49}, 238 (1994)
  [arXiv:hep-ph/9307290].


\bibitem{Zhang:2010ur}
  J.~M.~Zhang and G.~L.~Wang,
  Chin.\ Phys.\ Lett.\  {\bf 27}, 051301 (2010)
  [arXiv:1003.5576 [hep-ph]].


\bibitem{Deandrea:1993ma}
  A.~Deandrea, N.~Di Bartolomeo, R.~Gatto and G.~Nardulli,
  Phys.\ Lett.\  B {\bf 318}, 549 (1993)
  [arXiv:hep-ph/9308210].

\bibitem{Li:2008ts}
  R.~H.~Li, C.~D.~Lu and H.~Zou,
  Phys.\ Rev.\  D {\bf 78}, 014018 (2008)
  [arXiv:0803.1073 [hep-ph]]; 
%
%
%
\bibitem{Li:2009xf}
  R.~H.~Li, X.~X.~Wang, A.~I.~Sanda and C.~D.~Lu,
  Phys.\ Rev.\  D {\bf 81}, 034006 (2010)
  [arXiv:0910.1424 [hep-ph]].






\end{thebibliography}
\end{document}